\documentclass[nofootinbib]{revtex4}
\usepackage{graphicx}
\usepackage{dcolumn}
\usepackage{amsmath,amssymb,epsfig}
\usepackage{paralist}
\usepackage{comment}
\usepackage{graphicx}
\usepackage{multirow}
\usepackage{color,soul}
\usepackage{suffix}
\usepackage{mathtools}

\renewcommand{\vec}[1]{\boldsymbol{\mathrm{#1}}}

\begin{document}

\title{Navigating stellar wobbles for imaging with the solar gravitational lens}

\author{Slava G. Turyshev$^{1}$, Viktor T. Toth$^2$}

\affiliation{\vskip 3pt$^1$Jet Propulsion Laboratory, California Institute of Technology,\\
4800 Oak Grove Drive, Pasadena, CA 91109-0899, USA}

\affiliation{\vskip 3pt
$^2$Ottawa, Ontario K1N 9H5, Canada}

\date{\today}

\begin{abstract}

The solar gravitational lens (SGL) offers unique capabilities for direct high-resolution imaging of faint, distant objects, such as exoplanets. For that purpose, in the near future, a spacecraft carrying a meter-class telescope with a solar coronagraph would be placed in the focal region of the SGL. That region begins at $\sim547$~astronomical units from the Sun and occupies the immediate vicinity of the target-specific primary optical axis -- the line that connects the center of the target and that of the Sun. Clearly, this axis is not at rest. It undergoes complex motion as the exoplanet orbits its host star, as that star moves with respect to the Sun, and even as the Sun itself moves with respect to the solar system's barycenter due to the gravitational pull of planets in our solar system. Although less prominent, other motions exist.  An image of an extended object is projected by the SGL into an image plane and moves within that plane, responding to the motion of the optical axis. To sample the image, a telescope must always be on the move, following the projection, with precise knowledge of its own position with respect to the image. We consider the dominant motions that determine the position of the focal line as a function of time. We evaluate the needed navigational capability for the telescope to conduct a multiyear exoplanet imaging mission in the focal region for the SGL. We show that even in a rather conservative case, when an Earth-like exoplanet is in our immediate stellar neighborhood at $\sim10$ light years, the motion of the image is characterized by a small total acceleration that is driven primarily by the orbital motion of the exoplanet (its effect on the projected image estimated to be at the level of $\sim 6\,\mu {\rm m/s}^2$, decreasing inversely with distance to a target) and by the reflex motion of our Sun (target independent, contributing at $\sim 0.2\,\mu {\rm m/s}^2$). We discuss how the amplified light of the host star allows establishing a local reference frame that significantly relaxes navigational requirements for the imaging operations.
We conclude that the required navigation in the SGL's focal region, although complex, can be accurately modeled and a $\sim 10$-year prospective imaging mission is achievable with the already available propulsion technology.

\end{abstract}

\maketitle

\section{Introduction}
\label{sec:aintro}

According to Einstein's theory of general relativity, gravitation induces refractive properties on spacetime. Gravitationally deflected rays of light that pass on opposing sides of a mass converge. As a result, a massive object acts as a lens by bending incident photon trajectories inward. The resulting gravitational lensing is widely used in astronomy today to explore the Universe, including attempts to discover exoplanets.

As rays of light that pass farther from the lensing mass are deflected by smaller angles, a gravitational lens does not have a single focal point. Rather, it acts as a lens with substantial negative spherical aberration, characterized by a focal half-line. Of the solar system bodies, only the Sun is massive enough to have its focal half-line begin at a distance, $\sim 547$ astronomical units (AU), that is within the range of a realistic mission. This is the distance that corresponds to rays of light with the smallest impact parameter possible, i.e., light rays that graze the solar limb. The focal region extends far beyond 1,500 AU \cite{Turyshev:2017,Turyshev-Toth:2017}. With its significant light amplification and angular resolution of $\sim10^{11}$ and 0.1 nano-arcseconds at the wavelength of $\lambda=1\,\mu$m, \cite{Turyshev-Toth:2020-extend}, the SGL offers  unique capabilities for observing exoplanets or other faint, distant sources, such as regions in the vicinity of the super massive black holes, QSOs, etc.

All previous work on the SGL (and, in fact, on gravitational lensing, in general) assumed a static configuration with neither the source nor the lens moving.  That assumption was justified as it allowed us to develop new analytical tools, to establish the optical properties of the SGL, and to explore SGL's imaging capabilities by developing a comprehensive wave-optical approach to describe imaging with the SGL \cite{Turyshev-Toth:2017,Turyshev-Toth:2019-extend,Turyshev-Toth:2019-blur,Turyshev-Toth:2019-image,Turyshev-Toth:2020-extend,Turyshev-Toth:2021-multipoles}.

Clearly, a realistic imaging configuration is not static. For any observed source, its primary optical axis---the line connecting the center of the source and that of the Sun---moves. When the source is an exoplanet, the largest effects are due to its orbital motion around its host star, the motion of the host star with respect to the Sun and the motion of the Sun with respect to the solar system barycenter (i.e., the reflex motion under the gravitational pull from the planets, mostly gaseous giants.) Although much smaller, other motions exist. These motions determine the image position with respect to the Sun, its velocity and its acceleration.

Understanding the motion of the optical axis associated with an exoplanet is critical to the success of any prospective exoplanet imaging mission utilizing the SGL. Light from an exoplanet is projected to an image that is typically several kilometers in size in the image plane, a hypothetical plane that is perpendicular to the optical axis and contains the imaging instrument. The instrument, an observing telescope that views the Einstein ring around the Sun formed by light from the exoplanet, must measure the intensity of this light from many different locations within the image area. These measurements may require lengthy integration times, to compensate for the weakness of the exoplanetary signal received on the noisy background of the solar corona (i.e., to improve the signal-to-noise ratio). During these integration intervals, the instrument must remain at rest relative to the projected exoplanet image, which by itself moves within the image plane. Between measurements, the change in the imaging instrument's position with respect to the exoplanet image must again be well known in order to understand which part of the image is being measured, information essential to image reconstruction (see discussion in \cite{Turyshev-Toth:2020-extend,Toth-Turyshev:2020}).

Therefore, to carry out a successful imaging campaign, it is crucial to know precisely the position and, especially, the velocity of the imaging instrument with respect to the exoplanet's optical axis in the image plane. The related total velocity change (i.e., $\Delta v$) determines the quantity of onboard fuel and power needed to perform the mission.
Recognition of this need is our motivation for our detailed analysis. Our paper represents a first effort to systematically account for all major contributions that determine the motion of an exoplanet's image, as projected by the SGL, in an image plane that is itself changing as a result of the observing telescope's egress from the solar system \cite{Turyshev-etal:2020-PhaseII}.

As shown in \cite{Turyshev-Toth:2013,Turyshev:2012nw,Turyshev-GRACE-FO:2014}, the ultimate observational model for an SGL imaging mission may need to include relativistic terms and a typical set of small forces acting on the spacecraft, as discussed in \cite{Moyer:1971,Moyer:1981,Turyshev-Toth:2010LRR}.
This is needed if spacecraft navigation is done from the solar system barycentric coordinate reference frame (BCRF) --  the reference frame typically used to navigate the interplanetary spacecraft. However, as we shall see below, SGL navigation will be done using a local coordinate reference frame that will be established by relying on light from the host star and amplified by the SGL. (The availability of such a local reference frame, as discussed here, is mission-enabling.) But first, we must address the effects of the largest magnitude that will impact actual mission operations. This is needed in order to evaluate the required navigational precision and to develop a path toward feasible mission requirements. The goal is to establish sensible limits on the cumulative velocity change for an observing telescope that samples the exoplanet's projected image over an extended period of time, as well as the maximum acceleration that a spacecraft carrying such a telescope will need to facilitate. These results will directly contribute to an SGL mission design.

Here we consider only the largest contributions to the temporally changing image position, representing the \emph{coarse navigation objective} for the imaging telescope. Actually locating the exoplanet's projected image once the general vicinity is known (being guided by the amplified light from the host star) is the \emph{medium navigational objective}. Sampling the exoplanet image in a pixel-by-pixel fashion will be done by moving spacecraft within the image in a controlled pattern thus establishing the \emph{fine navigational objective}, which is also being considered but will be discussed elsewhere.

This paper is organized as follows:
In Section~\ref{sec:real-meas}, we introduce the coordinate reference frames that are used in the imaging applications with the SGL. We discuss our model of those dynamical effects that dominate the motion of the optical axes of the exoplanet and its host star in the image plane. We present the characteristics of the model exoplanetary system that is used in this investigation.
In Section~\ref{sec:position}, we consider the largest effects present in the positional displacements of the images, the optical axis of the host star and that of the target exoplanet.
In Section~\ref{sec:im-vel-acc} we discuss the corresponding velocities and accelerations.  We show that, even in the most demanding cases, the magnitudes of all dominant dynamical effects remain well within the range that can be addressed by readily available technical capabilities.
In Section \ref{sec:RF-and-imaging} we discuss the practical steps toward imaging with the SGL. We address the anticipated knowledge on the exoplanet prior to mission launch and outline the approach to establish local image-centric reference frames.
In Section \ref{sec:disc}, we summarize the results and discuss next steps.
To streamline the discussion, we moved some material to Appendices.
Appendix \ref{sec:Kepler} that provides a set of Keplerian expressions needed to model the position, velocity and acceleration for a given set of orbital elements.

\section{Toward realistic measurements}
\label{sec:real-meas}

We begin our investigation by considering an extended source at a finite distance, $z_0$, from the Sun. We use a cylindrical coordinate system $(\rho,z,\phi)$, with the $z$-axis corresponding to the preferred axis, defined as a line that connects the center of the source with the center of the Sun. Furthermore, we characterize points in the image plane and the source plane (both perpendicular to the $z$-axis) using 2-dimensional vector coordinates $\vec{x}$ and $\vec{x}'$, respectively, with the origin defined by the intersection of the optical axis with the respective planes.

We model light as a monochromatic high-frequency EM wave (i.e., neglecting terms $\propto(kr)^{-1}$, where $k=2\pi/\lambda$ is the wavenumber and $\lambda$ is the wavelength), arriving from a source at the distance of $z_0\gg r_g$, where $r_g=2GM/c^2$ is the Schwarzschild radius of the Sun. The wave is assumed to pass by the Sun and, depending on a particular light ray's impact parameter $b$, converges on the opposite side of the Sun at the heliocentric distance $ z =b^2/2r_g=547\, (b/R_\odot)^2$~AU. The corresponding EM field near the optical axis in the strong interference region of the SGL is given, up to ${\cal O}(\rho^2/z^2, \sqrt{2r_g z}/z_0)$, by the following expression \cite{Turyshev-Toth:2019-extend,Turyshev-Toth:2019-blur,Turyshev-Toth:2019-image}:
{}
\begin{eqnarray}
    \left( \begin{aligned}
{E}_\rho& \\
{H}_\rho& \\
  \end{aligned} \right) =    \left( \begin{aligned}
{H}_\phi& \\
-{E}_\phi& \\
  \end{aligned} \right)&=&
  \frac{E_0}{z_0}  \sqrt{2\pi kr_g}
  J_0\Big(k
\sqrt{{2r_g  z}}\,
\Big|\frac{{\vec x}}{  z}+\frac{{\vec x'}}{{ z}_0}\Big|\Big)
    e^{i\Omega(t)}
 \left( \begin{aligned}
 \cos\phi& \\
 \sin\phi& \\
  \end{aligned} \right),
  \label{eq:DB-sol-rho}
\end{eqnarray}
where the $z$-components of the EM wave are negligible behaving as $({E}_z, {H}_z)\sim {\cal O}({\rho}/{z}, \sqrt{2r_g z}/z_0)$, and the phase is given $\Omega(t)=k(r+r_0+r_g\ln [2k(r+r_0)])+\sigma_0-\omega t$, and where $J_0(x)$ is the Bessel function of the first kind \cite{Abramovitz-Stegun:1965}.

We use the solution (\ref{eq:DB-sol-rho}) and study the Poynting vector, ${\vec S}=(c/4\pi)\big<\overline{[{\rm Re}{\vec E}\times {\rm Re}{\vec H}]}\big>$, that describes the energy flux in the image plane \cite{Wolf-Gabor:1959,Richards-Wolf:1959,Born-Wolf:1999}. Normalizing this flux to the time-averaged value that would be observed if the gravitational field of the Sun were absent, $|\overline{\vec S}_0|=(c/8\pi)E_0^2/z_0^2$, we define the amplification factor of the SGL, ${ \mu}_{\tt SGL}=|{\vec S}|/|\overline{\vec S}_0|$:
{}
\begin{eqnarray}
{ \mu}_{\tt SGL}({\vec x},{\vec x}')&=&
\mu_0 \cdot {\rm PSF}({\vec x}, {\vec x}')
\qquad {\rm with} \qquad
\mu_0=2\pi kr_g\simeq1.17\times 10^{11}\,
\Big(\frac{1\,\mu{\rm m}}{\lambda}\Big),
\label{eq:S_z*6z-mu2}
\end{eqnarray}
where ${\rm PSF}({\vec x}, {\vec x}')$ is the point-spread function (PSF) of the monopole SGL:
{}
\begin{eqnarray}
{\rm PSF}({\vec x},{\vec x}')&=&
J^2_0\Big(k \sqrt{{2r_g  z}}\,
\Big|\frac{{\vec x}}{  z}+\frac{{\vec x'}}{{ z}_0}\Big|\Big).
\label{eq:psf}
\end{eqnarray}

For a given target and the corresponding primary optical axis, the PSF sets the geometry of both the source and image planes. In particular, the argument of the PSF implies the existence of the following scaling relation between the positions on the source plane, $\vec x'$, and those on the  image plane, $\vec x$:
 {}
\begin{eqnarray}
{\vec x}=-\frac{ z }{{ z}_0}{\vec x'} \qquad \Rightarrow \qquad
{\vec x} =1.03\times 10^{-3}\Big(\frac{ z }{650~{\rm AU}}\Big)\Big(\frac{10~{\rm ly} }{{ z}_0}\Big) {\vec x}'.
\label{eq:S_z*6z-pos}
\end{eqnarray}
We will use this relation to evaluate the needed precision of various quantities involved in the imaging observations.

To image a preselected target, we assume a meter-class telescope (equipped with an internal coronagraph needed to block the light from our own Sun as well as its inner corona region), placed within the target's image that is formed in the focal region of the SGL \cite{Turyshev-etal:2018,Turyshev-etal:2020-PhaseII}.  As the telescope egresses from the solar system with velocity $v_{\tt sc}$, its heliocentric distance increases with time as $z(t)=547\,{\rm AU}+v_{\tt sc}\cdot t$. Therefore, according to (\ref{eq:S_z*6z-pos}), for a spherical and fully illuminated source with diameter $d_{\tt source}$, the image of that object is projected by the SGL in a 3-D volume whose shape is a truncated cone with the diameter slowly increasing outward as $d_{\tt image}(t)\simeq(z(t)/z_0)d_{\tt source}$.

For imaging purposes, we define a set of image planes at various heliocentric distances in the SGL focal region and positioned perpendicular to the focal line. Imaging of a distant target is accomplished by placing a telescope in the focal region and moving it within the image volume in a pixel-by-pixel fashion while carefully determining its position within the image volume at any given moment. We assume that the telescope is following the outward leg of a hyperbolic heliocentric trajectory, therefore over time, it moves outward in the focal region, sampling the same image of slowly increasing size that is being projected into progressive image planes.

From an instantaneous position in a particular image plane, the telescope looks back at the Sun and observes the Einstein ring around the Sun that is formed by the SGL, amplifying light received from the target. The overall brightness of the Einstein ring is different at each image pixel location, representing the true scene variability of the source that is convolved with the SGL PSF. By moving the telescope within the image in a pixel-by-pixel fashion, the telescope collects brightness data at each pixel. These observations are combined to form an image of the target as projected by the SGL. A true image of the target is recovered by the process of deconvolution (applying the inverse of the SGL PSF) while also evaluating sources of noise, and possibly applying image reconstruction and noise reduction techniques to improve the signal-to-noise ratio and recover the image of the target at the best possible quality \cite{Turyshev-Toth:2020-extend,Toth-Turyshev:2020}.

\subsection{Coordinate reference frames and position vectors}
\label{sec:definition}

\begin{figure}
\includegraphics[width=0.70\linewidth]{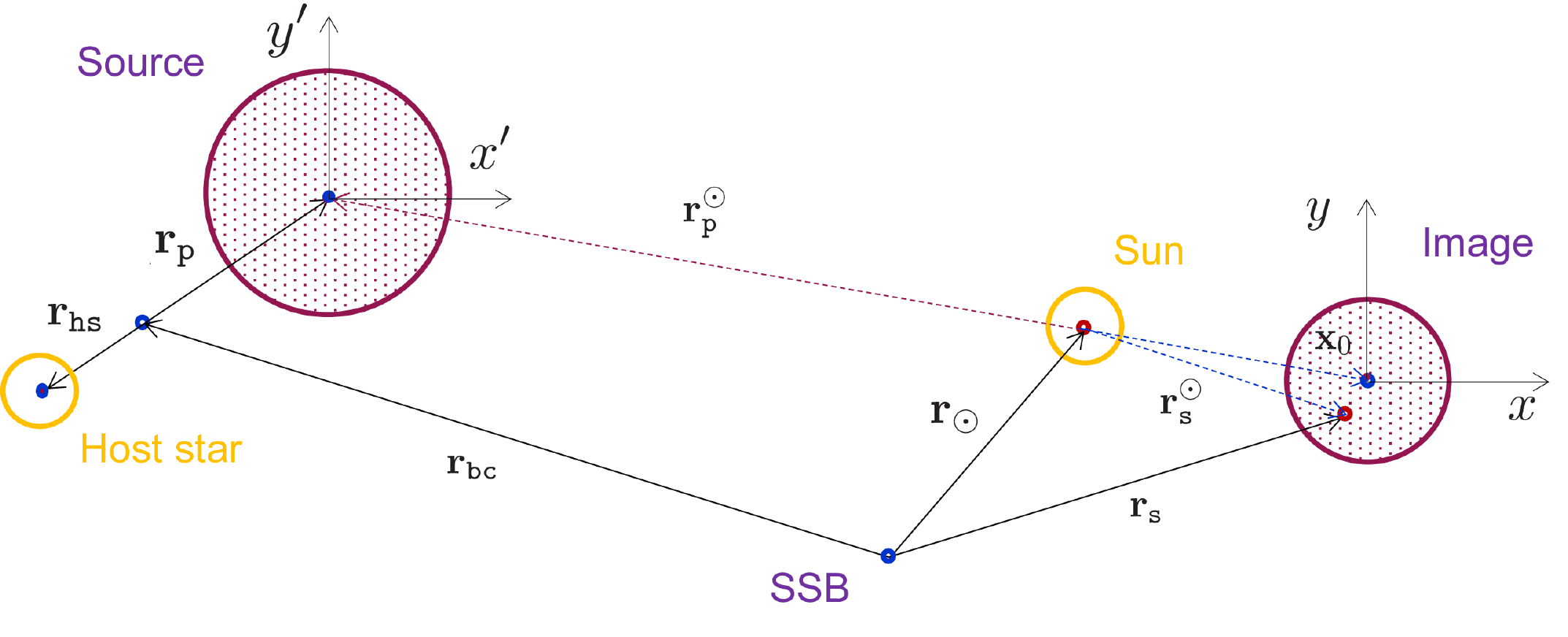}
\caption{\label{fig:SGN-model} Coordinate frames and the relevant position vectors involved in using the SGL for imaging purposes.
}
\end{figure}

To assess the navigational precision that is required for high-resolution imaging of an exoplanet with the SGL, we need to consider the relevant observable quantities. For that, we need to introduce several coordinate reference systems \cite{Turyshev-Toth:2013} that will help to describe the time-varying imaging geometry. The set of typical frames include those that are associated with celestial bodies (or planetary system), such as the following five frames (see Fig.~\ref{fig:SGN-model}):
\begin{itemize}
\item The solar system barycentric (SSB) coordinate reference frame. The SSB is a quasi-inertial reference frame and, as such, it is    considered to be the primary reference frame to describe the motion of celestial bodies of the solar system and interplanetary spacecraft \cite{Moyer:1971,Moyer:1981}. The SSB is not associated with any physical body, but is defined (i.e., pre-computed) using data from interplanetary spacecraft and astronomical observations.
\item The heliocentric coordinate reference frame. The origin of this frame is  associated with the center of gravity of the Sun. However, the Sun moves with respect to the SSB under the combined gravitational pull from the planets of the solar system (dominated by the gas giants, Jupiter and Saturn in particular). The heliocentric frame is noninertial. Nevertheless, this frame is the principal frame that is used to describe the SGL imaging.
\item The host star planetary system's barycentric coordinate reference frame. Similar to the SSB, this frame is quasi-inertial
and is the most convenient frame to describe the internal dynamics of the exoplanetary system.
\item The host star's coordinate reference frame. Similarly to the heliocentric frame, the origin of this frame is associated with the host star's center of mass. This frame is noninertial and may exhibit complex reflex motion \cite{Perryman:2011}, which must be accounted for during imaging operations.
\item The planet-centric coordinate reference frame. The origin of this frame is associated with the center of inertia of the exoplanet. This frame is also noninertial, which must be accounted in science operations.
\end{itemize}

These five coordinate reference frames are typically used to describe dynamics of gravitationally bound systems. They are defined and maintained using well-established procedures and are enabled by various astronomical measurements  and techniques \cite{Moyer:1971,Moyer:1981,TMT:2011,Turyshev:2012nw}.
For imaging with the SGL, however, we need to introduce new references frames that are not associated with a physical object but rather, use a light field for the reference purposes. In particular,  we need to establish image-centric coordinates that rely on the source's light amplified by the SGL and projected into an image plane in the SGL focal region. Specifically, the following two frames are of great importance:
 {}
\begin{itemize}
\item The host star image-centric coordinate reference frame.  As light from the host star is greatly amplified by the SGL, the  Einstein ring corresponding to such a self-luminous object will be much brighter than light from the solar corona. Such a bright Einstein ring allows for a reliable construction of the local reference frame in the focal region of the SGL. This noninertial frame is precomputed using the heliocentric position of the host star  \cite{Turyshev-Toth:2019-extend}. This frame is a scaled, inverted image of the host star centric frame (discussed above), with all the position vectors scaled by the ratio $z/z_0$, according to (\ref{eq:S_z*6z-pos}).
\item The exoplanet image-centric coordinate reference frame.  This frame is not associated with any celestial body and is established by using data from the image sensor of the observing telescope.  This frame is precomputed using the known orbit of the target exoplanet \cite{Turyshev-Toth:2019-extend}. Similarly to the host star image-centric frame, this frame is noninertial and is a scaled, inverted  image of the planet-centric reference frame, with position vectors scaled by the ratio $z/z_0$, in accord with (\ref{eq:S_z*6z-pos}). Establishing this reference frame is a primary prerequisite for exoplanet imaging operations. This can be done using the image sensor data on the imaging telescope and by precision tracking, using the previously established host star image-centric reference frame.
\end{itemize}

In Section~\ref{sec:loc-RefFrames} we describe a way to construct these reference frames using the image sensor of the observing telescope. Fig.~\ref{fig:SGN-model} shows all the coordinate reference frames and introduces the corresponding position vectors. These sets of coordinates allow us to introduce position vectors that are either known, i.e., directly observed via a certain type astronomical measurements, or are computed, i.e., those that are not directly observed, but are derived from the known position vectors. Specifically:
\begin{itemize}
\item Known position vectors:\\[-12pt]
\begin{itemize}
\renewcommand{\labelitemii}{\tiny{$\blacklozenge$}}
\item $\vec r_\odot=r_\odot \vec n_\odot$, is the position vector of the Sun in the SSB frame. This vector is known from tracking interplanetary spacecraft and from astronomical observations (primarily VLBI data). This displacement represents the reflex motion of the Sun under the influence of the planets of the solar system. Its magnitude may be larger than the solar radius. Its dynamics comes with different frequencies representing the various planetary influences on the Sun, especially those of the giant planets \cite{Perryman:2011}.
\item $\vec r_{\tt bc}= r_{\tt bc}\vec n_{\tt bc}$, is the position of the host star system's barycenter. This vector is assumed to be known from astronomical observations (e.g., using radial velocity, astrometry and parallax data).
\item $\vec r_{\tt hs}= r_{\tt hs}\vec n_{\tt hs}$, is the position of the host star with respect to the SSB. This vector is assumed known from astronomical observations (e.g., using radial velocity, astrometry and parallax data).
\item $\vec r_{\tt p}= r_{\tt p}\vec n_{\tt p}$, is the position of the exoplanetary target  with respect to its host star.  This vector is assumed known from astronomical data (e.g., using the data obtained from either planetary transits, astrometry, radial velocity, etc.)
\item $\vec r_{\tt s}= r_{\tt s}\vec n_{\tt s}$, is the position vector of an observing spacecraft in the strong interference region of the SGL.  This vector is known from spacecraft tracking data (either radio or optical ranging, Doppler and/or astrometry.)
\end{itemize}
\item Computed position vectors:\\[-12pt]
\begin{itemize}
\renewcommand{\labelitemii}{\tiny{$\blacklozenge$}}
\item $\vec r_{\tt p}^\odot= r_{\tt p}^\odot\vec n_{\tt p}^\odot$, is the heliocentric position of the exoplanetary target. This vector is determined from a combination of the following three known vectors:
\begin{equation}
\vec r_{\tt p}^\odot= \vec r_{\tt bc}+ \vec r_{\tt p}-\vec r_\odot.
\label{eq:vec_p}
\end{equation}

Note that vector $\vec r_{\tt p}^\odot$ establishes the direction of the SGL exoplanet primary optical axis, $\vec n_0$, which originates at the Sun and continues behind the Sun in the direction opposite to $\vec n_{\tt p}^\odot$, namely $\vec n_0=-\vec n_{\tt p}^\odot$. Vector $\vec n_0$ is used to establish  the exoplanet image-centric coordinate frame for a given  heliocentric imaging distance.

\item $\vec r_{\tt hs}^\odot= r_{\tt hs}^\odot\vec n_{\tt hs}^\odot$, is the heliocentric position of the host star. Similarly to (\ref{eq:vec_p}), this vector is determined from a combination of the following three known vectors:
\begin{equation}
\vec r_{\tt hs}^\odot= \vec r_{\tt bc}+ \vec r_{\tt hs}-\vec r_\odot.
\label{eq:vec_hs}
\end{equation}

Similarly, the SGL's host star primary optical axis, $\vec n_\star$,  is given as $\vec n_\star=-\vec n_{\tt hs}^\odot$. Vector  $\vec n_\star$ provides the key information to establish the host star image-centric coordinate reference frame.

\item $\vec r_{\tt s}^\odot= r_{\tt s}^\odot \vec n_{\tt s}^\odot$, is the heliocentric position of the observing spacecraft. This vector is determined from a combination of the following two known vectors:
\begin{equation}
\vec r_{\tt s}^\odot=  \vec r_{\tt s}-\vec r_\odot.
\label{eq:vec_s}
\end{equation}
For imaging purposes, we place telescope within a particular image (i.e., either that of the exoplanet(s) or the host star) and thus determine the telescope's position with respect to a particular primary optical axis, either $\vec n_0$ or $\vec n_\star$. In this case, we need to achieve the highest accuracy in the determining  components of (\ref{eq:vec_s}) in the direction transverse to $\vec n_0$ (or $\vec n_\star$), i.e., within the image plane.
\end{itemize}
\end{itemize}

Other vectors may be introduced when needed, especially to describe imaging operations.

\subsection{Temporal variability in the image position}
\label{sec:mod-prim-opt-ax}

\subsubsection{Realistic modeling of the imaging geometry}

To describe realistic imaging observations with the SGL, we need to consider three sources of important dynamics: the mutual positions of the source and the lens, the overall imaging geometry, and the heliocentric position of the imaging telescope. The mutual positions establish the primary optical axis. The imaging geometry specifies the image formed by the SGL in its  focal region. The telescope's heliocentric position gives pixel position where the light field that represents the image is observed. We will use this logic to develop the relevant observational model.

We begin by defining the primary optical axis for an exoplanet that we denote $\vec n_0$. Given the overall imaging geometry (shown in Fig.~\ref{fig:SGN-model}), the primary optical axis is defined by the unit vector that points from the exoplanet's center of gravity toward that of the Sun. Using (\ref{eq:vec_p}), we determine this vector as  $\vec n_0=-{\vec r_{\tt p}^\odot}/{r_{\tt p}^\odot}$. Taking into account that $r_{\tt p}, r_\odot \ll r_{\tt bc}$, we can express the direction of the exoplanet primary optical axis via known position vectors, which, essentially, results in the following heliocentric vector:
{}
\begin{equation}
\vec n_0=-\vec n_{\tt p}^\odot=-\frac{\vec r_{\tt p}^\odot}{r_{\tt p}^\odot}= -\frac{\vec r_{\tt bc}+ \vec r_{\tt p}-\vec r_\odot}{|\vec r_{\tt bc}+ \vec r_{\tt p}-\vec r_\odot|}=-\Big\{\vec n_{\tt bc}+\frac{1}{r_{\tt bc}}\big[\vec n_{\tt bc}\times\big[ ({\vec r_{\tt p}-\vec r_\odot})\times\vec n_{\tt bc}\big]\big]+{\cal O}\Big(\frac{r^2_{\tt p}}{r^2_{\tt bc}},\frac{r^2_\odot}{r^2_{\tt bc}}\Big)\Big\},
\label{eq:vec_popt*}
\end{equation}
where $r=|\vec r|$ and $[\vec a \times \vec b]$ stands for the vector product of  two vectors $\vec a$ and $\vec b$.
We will use this expression to evaluate the magnitudes of all the vectors involved and also to establish the relevant mission requirements.

Next we consider an image plane, positioned at heliocentric distance $d_{\tt p}$ in the focal region of the SGL and oriented perpendicular to the primary optical axis, ${\vec n}_0$. We take an image pixel with coordinates ${\vec x}$ on that plane. The SSB position of that pixel  is given as
{}
\begin{equation}
{\vec r}^{\tt im}_{\tt p}={\vec r}_\odot+d_{\tt p}{\vec n}_0+{\vec x},
\label{eq:vec_rim-p}
\end{equation}
where the last two terms represent the heliocentric part of the vector and the first term is the SSB shift.

To conduct imaging observations, we position an imaging telescope in that image pixel. Then, using the heliocentric position of the telescope, ${\vec r}_{\tt s}^\odot$, introduced by (\ref{eq:vec_s}),  its SSB position is given as
{}
\begin{equation}
\vec r_{\tt s}={\vec r}_\odot+({\vec r}_{\tt s}^\odot\cdot{\vec n}_0){\vec n}_0+[{\vec n}_0\times[{\vec r}_{\tt s}^\odot\times {\vec n}_0]],
\label{eq:r_sc}
\end{equation}
where, again, the last two terms represent the heliocentric position and the first term is the SSB shift.

Results (\ref{eq:vec_rim-p}) and (\ref{eq:r_sc}) allow us to establish relationships between the quantities that describe the overall imaging geometry. By equating these expressions, $r^{\tt im}_{\tt p}=\vec r_{\tt s}$, we see that the heliocentric distance to the image plane, $d_{\tt p}$, and an image pixel position on that plane, ${\vec x}$, are provided by the heliocentric telescope position:
{}
\begin{eqnarray}
d_{\tt p}&=&({\vec r}_{\tt s}^\odot\cdot{\vec n}_0), \qquad
{\vec x}=[{\vec n}_0\times[{\vec r}_{\tt s}^\odot\times {\vec n}_0]].
\label{eq:vec_rim-dd}
\end{eqnarray}

Substituting these results in the scaling relation (\ref{eq:S_z*6z-pos}), we may now identify the source pixel position, $\vec x'$, in the source plane located at the heliocentric distance of $r_{\tt p}^\odot$,  in terms of the heliocentric position of the imaging telescope:
{}
\begin{eqnarray}
\frac{[{\vec n}_0\times[{\vec r}_{\tt s}^\odot\times {\vec n}_0]]}{({\vec r}_{\tt s}^\odot\cdot{\vec n}_0)}=-\frac{{\vec x}'}{r_{\tt p}^\odot} \qquad\Rightarrow\qquad
\frac{[{\vec n}_0\times[{\vec n}_{\tt s}^\odot\times {\vec n}_0]]}{({\vec n}_{\tt s}^\odot\cdot{\vec n}_0)}=-\frac{{\vec x}'}{r_{\tt p}^\odot}.
\label{eq:vec_rim-sc}
\end{eqnarray}
This result may be used as input into the image reconstruction algorithms that are used to simulate realistic observing conditions and also to establish requirements on the navigational accuracy required for imaging.

Expression (\ref{eq:vec_rim-sc}) suggests that it is not the heliocentric distance of the telescope but its alignment with respect to the primary optical axis that plays the most important role for imaging.  Given that in the strong interference region $({\vec n}_{\tt s}^\odot\cdot{\vec n}_0)\simeq 1$, the imaging sensitivity to observe a particular source image pixel ${\vec x}'$ comes from the transverse components of ${\vec n}_{\tt s}^\odot$, transforming (\ref{eq:vec_rim-sc}) as
{}
\begin{eqnarray}
[{\vec n}_0\times[{\vec n}_{\tt s}^\odot\times {\vec n}_0]] \simeq-\frac{{\vec x}'}{r_{\tt p}^\odot}.
\label{eq:vec_rim-sc22}
\end{eqnarray}

The high-precision  alignment between the two vectors, ${\vec n}_0$ and ${\vec n}_{\tt s}^\odot$, implied by (\ref{eq:vec_rim-sc22}), depends on two factors:
\begin{inparaenum}[i)]
\item the quality of spacecraft navigation and
\item the pointing precision of the imaging telescope.
\end{inparaenum}
Expression (\ref{eq:vec_rim-sc22}) may be used to establish mission and instrument requirements on these factors. In the SSB or heliocentric frames, satisfying such an alignment may be challenging. However, this is not needed. The fact that we may establish the local reference frame relying on the amplified light of the host star (discussed above; to be further elaborated in Section~\ref{sec:RF-and-imaging}), greatly simplifies this critical element of mission design.

\subsubsection{Temporally varying position of the exoplanet image}
\label{sec:tem_exo}

In the previous subsection we were able to present the imaging geometry in terms of the direction of the  primary optical axes,  $\vec n_0=-{\vec r_{\tt p}^\odot}/{r_{\tt p}^\odot}$. As this vector is not directly observed, result (\ref{eq:vec_popt*}) allows us to express it via measurable quantities thus allowing us to evaluate  temporal behavior of $\vec n_0$. As we see, this vector is refereed to the barycenter of the exoplanetary system, $\vec n_{\tt bc}$, with deviations the transverse direction which is the image plane. We can now determine projections of the orbital motions on that image plane. As seen from (\ref{eq:vec_rim-dd}), to first order, the telescope samples the image plane of the exoplanet that coincides with the plane in the transverse direction to $\vec n_{\tt bc}$, namely
{}
\begin{equation}
\vec x=[{\vec n}_0\times[{\vec r}_{\tt s}^\odot\times {\vec n}_0]]=[{\vec n}_{\tt bc}\times[{\vec r}_{\tt s}\times {\vec n}_{\tt bc}]]\Big(1+{\cal O}(\frac{r_{\tt p}}{r_{\tt bc}},\frac{r_\odot}{r_{\tt bc}})\Big),
\label{eq:x_rs}
\end{equation}
where we accounted for the fact that $ r_\odot\ll  r_{\tt s}$. Thus, it will be instructive to present the quantities involved in their relations with respect to $\vec n_{\tt bc}$. For that, relying on the knowledge of the heliocentric distance $d_{\tt p}$ to the image plane  and the pixel position ${\vec x}$ in that plane, and using the fact that these quantities are given by the heliocentric position of the imaging telescope (\ref{eq:vec_rim-dd}), we can now consider (\ref{eq:vec_rim-p}) in more detail, emphasizing its temporal behavior.  Using ${\vec n}_0$  from (\ref{eq:vec_popt*}), we present (\ref{eq:vec_rim-p}) as
{}
\begin{eqnarray}
\vec r^{\tt im}_{\tt p}&=&
\vec r_\odot+\vec n_0 d_{\tt p}+{\vec x}=\nonumber\\
&=&
\Big((\vec r_\odot\cdot\vec n_{\tt bc})-d_{\tt p}\Big)\vec n_{\tt bc}+\big[\vec n_{\tt bc}\times\big[\vec r_\odot \times\vec n_{\tt bc}\big]\big]+
\frac{d_{\tt p}}{r_{\tt bc}}
\big[\vec n_{\tt bc}\times\big[ (\vec r_\odot}-{\vec r_{\tt p})\times\vec n_{\tt bc}\big]\big]+{\vec x}+{\cal O}\Big(\frac{r^2_{\tt p}}{r^2_{\tt bc}}, \frac{r^2_\odot}{r^2_{\tt bc}}\Big)d_{\tt p},
\label{eq:pl-pos-p2}
\end{eqnarray}
where ${\vec x}$ is from (\ref{eq:x_rs}). Also, given the fact that $ r_\odot\ll  r_{\tt s}$ and using (\ref{eq:vec_popt*}) and (\ref{eq:vec_rim-dd}), we may express $d_{\tt p}$ as
 {}
\begin{eqnarray}
d_{\tt p}&=&({\vec r}_{\tt s}^\odot\cdot{\vec n}_0)=
-\big((\vec r_{\tt s}-\vec r_\odot)\cdot\vec n_{\tt bc}\big)\Big(1+{\cal O}\Big(\frac{r^2_{\tt p}}{r^2_{\tt bc}}, \frac{r^2_\odot}{r^2_{\tt bc}}\Big)\Big).
\label{eq:ex-pos-im-dx2}
\end{eqnarray}

As a result, expression (\ref{eq:pl-pos-p2}) takes the form,
{}
\begin{eqnarray}
{\vec r}^{\tt im}_{\tt p}&=& (\vec r_{\tt s}\cdot\vec n_{\tt bc})\vec n_{\tt bc}+{\vec r}^\perp_{\tt p 0}+[{\vec n}_{\tt bc}\times[{\vec r}_{\tt s}^\odot\times {\vec n}_{\tt bc}]]
+{\cal O}\big(\frac{r^2_{\tt p}}{r^2_{\tt bc}}, \frac{r^2_\odot}{r^2_{\tt bc}}\big)r_{\tt s}.
\label{eq:pl-pos-p2+}
\end{eqnarray}
The first term in (\ref{eq:pl-pos-p2+}) is the distance to the image plane that is evaluated along $\vec n_{\tt bc}$. The second term, ${\vec r}^\perp_{\tt p 0}$, is the position of the primary optical axis on the image plane that is given as
{}
\begin{eqnarray}
{\vec r}^\perp_{\tt p 0}&=&
\big[\vec n_{\tt bc}\times\big[\vec r_\odot \times\vec n_{\tt bc}\big]\big]\Big(1-\frac{(\vec r_{\tt s}\cdot\vec n_{\tt bc})}{r_{\tt bc}}\Big)+
\frac{(\vec r_{\tt s}\cdot\vec n_{\tt bc})}{r_{\tt bc}}
\big[\vec n_{\tt bc}\times\big[\vec r_{\tt p}\times\vec n_{\tt bc}\big]\big].
\label{eq:pl-pos-p2+perp}
\end{eqnarray}
Finally, the third term is the position of the imaging telescope within the image of the exoplanet in the image plane.

\subsubsection{Temporally-varying position of the host star image}

Relationships similar to (\ref{eq:pl-pos-p2+perp}) may be established to describe the image of the host star. We simply replace $\vec n_0$ with the unit vector in the direction of the primary optical axis of the host star, $\vec n_\star=-{\vec r_{\tt hs}^\odot}/{r_{\tt hs}^\odot}$, which in the heliocentric frame is given as
{}
\begin{equation}
\vec n_\star=-\vec n_{\tt hs}^\odot=-\frac{\vec r_{\tt hs}^\odot}{r_{\tt hs}^\odot}= -\frac{\vec r_{\tt bc}+ \vec r_{\tt hs}-\vec r_\odot}{|\vec r_{\tt bc}+ \vec r_{\tt hs}-\vec r_\odot|}=-\Big\{\vec n_{\tt bc}+\frac{1}{r_{\tt bc}}\big[\vec n_{\tt bc}\times\big[ ({\vec r_{\tt hs}-\vec r_\odot})\times\vec n_{\tt bc}\big]\big]+{\cal O}\Big(\frac{r^2_{\tt hs}}{r^2_{\tt bc}},\frac{r^2_\odot}{r^2_{\tt bc}}\Big)\Big\}.
\label{eq:vec_popt_hs*}
\end{equation}

To describe imaging observations of the host star, we introduce the SSB position of the telescope used to image the host star, $\vec r_{\tt s}^{\tt hs}$; so, in general, $\vec r_{\tt s}^{\tt hs}\not= \vec r_{\tt s}$. Then, following the logic outlined in Sec.~\ref{sec:tem_exo} and taking into account the fact that $ r_\odot\ll  r^{\tt hs}_{\tt s}$, we determine the distance to the  host star image plane, $d_{\tt hs}$, and position of the telescope within the host star image, ${\vec x}_{\tt hs}$, are given as
\begin{eqnarray}
d_{\tt hs}&=&-\big((\vec r^{\tt hs}_{\tt s}-\vec r_\odot)\cdot\vec n_{\tt bc}\big)\Big(1+{\cal O}\Big(\frac{r^2_{\tt hs}}{r^2_{\tt bc}}, \frac{r^2_\odot}{r^2_{\tt bc}}\Big)\Big),  \\
{\vec x}_{\tt hs}&=&\big[\vec n_\star\times\big[(\vec r_{\tt s}^{\tt hs}-\vec r_\odot)\times\vec n_\star\big]\big]=
[{\vec n}_{\tt bc}\times[{\vec r}_{\tt s}^{\tt hs}\times {\vec n}_{\tt bc}]]\Big(1+{\cal O}(\frac{r_{\tt hs}}{r_{\tt bc}},\frac{r_\odot}{r_{\tt bc}})\Big).
\end{eqnarray}

As a result, the SSB position vector of a pixel on the host star image, $\vec r_{\tt hs}^{\tt im}$, may be given in the following form:
{}
\begin{eqnarray}
\vec r_{\tt hs}^{\tt im}&=&(\vec r_{\tt s}^{\tt hs}\cdot\vec n_{\tt bc})\vec n_{\tt bc}+ {\vec r}^\perp_{\tt hs 0}+ [{\vec n}_{\tt bc}\times[{\vec r}_{\tt s}^{\tt hs}\times {\vec n}_{\tt bc}]]
+{\cal O}\big(\frac{r^2_{\tt hs}}{r^2_{\tt bc}}, \frac{r^2_\odot}{r^2_{\tt bc}}\big)r^{\tt hs}_{\tt s},
\label{eq:hs-pos-p2}
\end{eqnarray}
where the first term is the SSB distance to the image plane, the second term, ${\vec r}^\perp_{\tt hs 0}$, is the SSB position of the host star's primary optical axis in the host star image plane, given as
{}
\begin{eqnarray}
{\vec r}^\perp_{\tt hs 0}&=&
\big[\vec n_{\tt bc}\times\big[\vec r_\odot \times\vec n_{\tt bc}\big]\big]\Big(1-\frac{(\vec r_{\tt s}^{\tt hs}\cdot\vec n_{\tt bc})}{r_{\tt bc}}\Big)+
\frac{(\vec r_{\tt s}^{\tt hs}\cdot\vec n_{\tt bc})}{r_{\tt bc}}
\big[\vec n_{\tt bc}\times\big[\vec r_{\tt hs}\times\vec n_{\tt bc}\big]\big],
\label{eq:hs-pos-p2-perp}
\end{eqnarray}
and the last term in (\ref{eq:hs-pos-p2}) is the telescope position in the image plane of the host star.

Eqs.~(\ref{eq:pl-pos-p2+perp}), (\ref{eq:hs-pos-p2-perp}) are the observable quantities characterizing exoplanet imaging and represent our primary concern.  Note that the first terms in  these expressions are nearly identical and are due to the solar reflex motion scaled by the effect of the egress motion of the telescope that represents the extra distance for light to travel before reaching the image plane. We observe that expressions (\ref{eq:pl-pos-p2+perp}) and (\ref{eq:hs-pos-p2-perp}) may be obtained from simple  geometrical considerations. (The ultimate dynamical model needed to estimate the real-time position of the primary optical axes on the image plane should include contributions of all small forces acting on the spacecraft, including Newtonian and general relativistic corrections as well as nongravitational forces, similar to those described in \cite{Turyshev-Toth:2010LRR,Turyshev-etal:2011-PA}.) Results (\ref{eq:pl-pos-p2+perp}) and (\ref{eq:hs-pos-p2-perp}) capture the largest effects influencing the dynamics of the primary optical axes under various motions. These expressions are well suited for the purposes of our investigation.

\subsection{Modeling imaging observations}
\label{sec:syst-mod}

To facilitate the analysis, we need to make assumptions about the dynamical parameters relevant for imaging with the SGL. Specifically, we need to capture the most prominent dynamical features of our solar system, model the exoplanetary system, and make some preliminary assumptions concerning mission design.

\subsubsection{Modeling the solar system and the host star system}

To represent our solar system, we model it as a group of planets that gravitationally interact only with the Sun while neglecting interactions between planets.  As we are interested in the evaluation of only the largest dynamical effects that may be used to inform the SGL mission design, we consider only the gas giants Jupiter, Saturn, Neptune and Uranus. We use well-known orbital parameters describing these planets, summarized in Table~\ref{tab:ssplanets}.

For the exoplanetary system, while we recognize that the systems of the greatest interest at present are systems like TRAPPIST-1\footnote{For information on the TRAPPIST-1 system, see \url{https://en.wikipedia.org/wiki/TRAPPIST-1.}}, we nonetheless opted to model a solar system more like our own. We assume that in the near future, new targets that are more similar to our own solar system will be discovered.
In addition, as we show below,  due to the resulting larger displacements in the SGL image plane, such a system represent an interesting navigational challenge for an SGL mission. Thus, using such a solar system analogue represents a rather conservative case.

Accordingly, we assume a host star that is identical to our Sun, located at the distance of $z_0=10$~ly ($\sim3.1$ pc) from us. Dynamical effects in the image plane are largest for nearby systems, so this choice represents somewhat of a worst-case scenario. We model the hypothetical exoplanetary system to be similar to our solar system, with four gas giants in addition to the Earth-like planet that is the target of our investigation. For the exoplanets, we assume the same values for the dynamical Keplerian parameters that characterize the gas giants in our solar system, namely $m^{\tt exo}_j=m_j$, $a^{\tt exo}_j=a_j$, $T^{\tt exo}_j=T_j$,  $e^{\tt exo}_j=e_j$, and  $t^{\tt exo}_{0j}=t_{0j}$, where $j=\{{\tt E, J, S, U, N}\}$ denote the set of the planets used in the analysis. However, we uniformly vary the angles characterizing their orbits, namely $\Omega^{\tt exo}_j= q\, \Omega_j$, $\omega^{\tt exo}_j= q\, \omega_j$ and $i^{\tt exo}_j= q\, i_j$, where $q$ is a constant parameter.

The introduction of the parameter $q$ allows us to study different orbital inclinations of an exo-Earth with respect to our solar system's ecliptic plane: it can be face-on (i.e., $q=1.0$, yielding $i^{\tt exo}_\oplus= i_\oplus$) or edge-on (i.e., $q=1.8\times 10^6$, yielding $i^{\tt exo}_\oplus=1.8\times 10^6\, i_\oplus\simeq90^\circ$). We chose $q=7.5\times10^5$, which corresponds to setting the inclination of the exo-Earth's orbit with respect to the solar system's ecliptic plane at $i^{\tt exo}_\oplus=7.5\times10^5\, i_\oplus\simeq 37.5^\circ$.

By introducing the parameter $q$ as a uniform scaling of all the angles involved, we effectively ``destroy'' the exoplanetary system's ecliptic plane (see Table~\ref{tab:ssplanets-exo}). The exoplanetary orbital inclinations now are vastly different and effectively randomized. By doing this, we allow for the presence of much stronger dynamical signatures in the reflex motion of the host star (i.e., its position, velocity, acceleration with respect to the barycenter of the host star's system). However, as we will show below, these effects being multiplied by the scaling relation (\ref{eq:S_z*6z-pos}) result in very small contributions in the SGL image plane. Alternatively, we studied the case of just adding $37.5^\circ$ to the inclinations of each of the planet in the modeled exoplanetary system. Such an approach resulted in even weaker dynamical signals in the motion of the image, which motivated us to consider the uniform scaling with the parameter $q=0.75\times10^6$, treating the resulting set of exoplanetary orbital parameters as a ``worst case'' scenario.

\begin{table*}[t!]
\vskip-15pt
\caption{Solar system planetary parameters used in the analysis: $m$ is the mass of a planet, $a$ is its semi-major axis, $T$ is its orbital period, $e$ is the eccentricity of its orbit, $\Omega$ is the ascending node longitude, $\omega$ is the argument of perihelion, $i$ is the inclination to the ecliptic plane and $t_0$ is the time of perihelion passage.
\label{tab:ssplanets}}
\begin{tabular}{|l|c|c|c|c|c|c|c|r|}\hline
Planet  & $m$, kg & $a$, AU &  $T$, yr& $e$ & $\Omega$, $[{}^\circ]$ & $\omega$, $[{}^\circ]$ &$i$, $[{}^\circ]$& \multicolumn{1}{|c|}{$t_0$}  \\\hline\hline
Earth& $5.97237\times 10^{24}$ &  1  & 1 & 0.0167086 & $\phantom{0}{-11.26064}$ & 114.20783\phantom{0} & 0.00005\phantom{0} &  Jan 2, 2021\\
\hline
Jupiter& $\phantom{0}1.8982\times 10^{27}$ & \phantom{0}5.2044  & $\phantom{-0}11.862\phantom{00}$ & 0.0489\phantom{000} &100.464 & 273.867\phantom{000}& 1.303\phantom{000} &  Jan 21, 2023\\
Saturn& $\phantom{0}5.6834\times 10^{26}$ & \phantom{0}9.5826  & $\phantom{-0}29.4571\phantom{0}$ & 0.0565\phantom{000}&113.665 & 339.392\phantom{000}& 2.485\phantom{000} &  Nov 29, 2032\\
Uranus& $\phantom{0}8.6810\times 10^{25}$ & 19.2184  & $\phantom{-0}84.0205\phantom{0}$ & 0.046381\phantom{0}& \phantom{0}74.006  & \phantom{0}96.998857 & 0.773\phantom{000} &  Aug 19, 2050\\
Neptune& $1.02413\times 10^{26}$ &  30.07\phantom{00}  & $\phantom{-}164.8\phantom{0000}$ & 0.008678\phantom{0} & 131.784 & 276.336\phantom{000} & 1.767957 &  Sep 4, 2042\\
\hline
\end{tabular}
\vskip 0pt
\caption{Fictitious exosolar system planetary parameters used in the analysis. Several orbital parameters that were used in the study, including the planetary masses $m$, their semi-major axes $a$,  orbital periods $T$, orbital eccentricities $e$,
and the times $t_0$ of perihelion passage are the same as in Table~\ref{tab:ssplanets}. However, all angles are uniformly scaled as $\Omega^{\tt exo}_j= q\, \Omega_j$, $\omega^{\tt exo}_j= q\, \omega_j$ and $i^{\tt exo}_j= q\, i_j$, where $q=0.75\times10^6$. The orbital velocities and accelerations of the planets are estimated as $v_j=(2\pi/T_j) a_j$ and $\dot v_j=(2\pi/T_j)^2 a_j$, correspondingly. The corresponding velocities and accelerations of the planetary images are scaled by the SGL in accordance with Eq.~(\ref{eq:S_z*6z-pos}): for $z_0=10$~ly and $z=650$~AU, the scaling yields $v^{\rm im}_j=1.03\times 10^{-3} v_j$ and $a^{\rm im}_j=1.03\times 10^{-3} \dot v_j$.
\label{tab:ssplanets-exo}}
\vskip -7pt
\begin{tabular}{|l|c|c|c||c|c||c|c|}\hline
Exoplanet  & $\Omega$, $[{}^\circ]$ & $\omega$, $[{}^\circ]$ &$i$, $[{}^\circ]$& $v_j$, [km/s]  & $\dot v_j$, [m/s${}^2$] & $v_j^{\rm im}$, [m/s] & $a_j^{\rm im}$, [m/s${}^2$] \\\hline\hline
Earth & $-240.00$ & $\phantom{-}352.50$ & \phantom{0}37.50 & 29.81 & $5.94\times 10^{-3}$ & 30.64 & $6.10\times 10^{-6}$\\
\hline
Jupiter& $\phantom{-00}0.00$ & $\phantom{-0}90.00$& 210.00 &  13.08&
$2.20\times 10^{-4}$ & 13.44&
$2.26\times 10^{-7}$\\
Saturn& $\phantom{-0}30.00$ & $-120.00$& \phantom{0}30.00& \phantom{0}9.70 &
$6.56\times 10^{-5}$& \phantom{0}9.97&
$6.74\times 10^{-8}$\\
Uranus& $\phantom{-0}60.00$  & $\phantom{0}{-17.25}$ & 150.00 & \phantom{0}6.82 &
$1.62\times 10^{-5}$& \phantom{0}7.01&
$1.66\times 10^{-8}$\\
Neptune& $\phantom{-00}0.00$ & $\phantom{-00}0.00$ & \phantom{0}87.75& \phantom{0}5.44 &
$6.58\times 10^{-6}$& \phantom{0}5.59&
$6.76\times 10^{-9}$\\
\hline
\end{tabular}
\end{table*}

Table~\ref{tab:ssplanets-exo} also provides estimates for the mean velocities and accelerations within the exoplanetary system as well as the same quantities for their images scaled by the SGL in accord to Eq.~(\ref{eq:S_z*6z-pos}).  These estimates already provide key information related to the accelerations of the exoplanetary images on the image plane. The largest acceleration of the image of the exo-Earth is $ 6.10\times 10^{-6}~(10~{\rm ly}/z_0)~{\rm m}/{\rm s}^2$. This characterizes the required acceleration for an SGL observing telescope in the image plane.

\subsubsection{Observing in the line of sight direction}
\label{sec:nbc-vec-k}

While results (\ref{eq:pl-pos-p2+perp}), (\ref{eq:hs-pos-p2-perp}) are generic and applicable to any planetary system, we need to make them a bit more convenient for our simulations. For that, we consider that the SSB positions of a telescope that images the exoplanet and another telescope that images the host star are nearly identical, $\vec r_{\tt s}^{\tt hs}\simeq \vec r_{\tt s}$. This assumption yields $d_{\tt p}\simeq d_{\tt hs}=-(\vec r_{\tt s}\cdot\vec n_{\tt bc})=-(\vec r^{\tt hs}_{\tt s}\cdot\vec n_{\tt bc})\equiv z$, where $z$ is the heliocentric distance to the imaging telescopes, present in (\ref{eq:S_z*6z-pos}). We also define $r_{\tt bc}=z_0$, where $z_0$ is the heliocentric target position, also present in (\ref{eq:S_z*6z-pos}). Next, without a loss of generality, we choose the exoplanetary system positioned in the direction perpendicular to the solar system ecliptic plane. We take this  direction to be the $z$-direction of the SSB coordinate system, namely, we choose $\vec n_{\tt bc}=\vec k$ and, thus, for any vector $\vec a$, the following is valid: $[\vec k\times[\vec a\times\vec k]]=a_x\vec e_x+a_y\vec e_y=\vec a^\perp$, with $\vec a^\perp$ being the component of $\vec a$ orthogonal to $\vec k$. Using this identity in (\ref{eq:pl-pos-p2+}) and (\ref{eq:hs-pos-p2}), we develop the following expressions that determine the SSB positions of the image pixels in the respective image planes of the exoplanet and the host star:
{}
\begin{eqnarray}
{\vec r}^{\tt im}_{\tt p}&=& z\,\vec k+{\vec r}^\perp_{\tt p 0}+
[{\vec k}\times[{\vec r}_{\tt s}^\odot\times {\vec k}]]
+{\cal O}\big(\frac{r^2_{\tt p}}{z^2_0}, \frac{r^2_\odot}{z^2_0}\big)z,
\label{eq:pl-pos-p2k+}\\
\vec r_{\tt hs}^{\tt im}&=&z\,\vec k+ {\vec r}^\perp_{\tt hs 0}+
[{\vec k}\times[{\vec r}_{\tt s}^{\tt hs}\times {\vec k}]]
+{\cal O}\big(\frac{r^2_{\tt hs}}{z^2_0}, \frac{r^2_\odot}{z^2_0}\big)z,
\label{eq:hs-pos-p2k}
\end{eqnarray}
with the positions of the primary optical axes in the image planes for the exoplanet, ${\vec r}^\perp_{\tt p 0}$,  and the host star, ${\vec r}^\perp_{\tt hs 0}$, from (\ref{eq:pl-pos-p2+perp}) and (\ref{eq:hs-pos-p2-perp}), given by
{}
\begin{eqnarray}
{\vec r}^\perp_{\tt p 0}&=&\Big(1+\frac{z}{z_0}\Big)\vec r^\perp_\odot-
\frac{z}{z_0}\vec r^\perp_{\tt p},
\label{eq:pl-pos-p2+perp-k}\\
{\vec r}^\perp_{\tt hs 0}&=&\Big(1+\frac{z}{z_0}\Big)\vec r^\perp_\odot-\frac{z}{z_0}\vec r^\perp_{\tt hs},
\label{eq:hs-pos-p2-perp-k}
\end{eqnarray}
where $\vec r^\perp_\odot=r_{\odot x}\vec e_x+ r_{\odot y}\vec e_y$, $\vec r^\perp_{\tt p}=  r_{{\tt p}\, x}\vec e_x+ r_{{\tt p}\, y} \vec e_y$ and $\vec r^\perp_{\tt hs}=r_{{\tt hs}\, x}\vec e_x+\vec r_{{\tt hs}\, y}\vec e_y$, are projections in the image plane of the solar reflex motion, exoplanetary and host star orbital position vectors, and $z\equiv z(t)$.

The first term in expressions (\ref{eq:pl-pos-p2+perp-k})--(\ref{eq:hs-pos-p2-perp-k}) are identical, resulting from the reflex motion of the Sun, scaled by the $1+z/z_0$ factor that is due to the extra heliocentric distance that light has to travel before it reaches the image plane. Their form may be understood from simple geometric reasoning: considering only the solar motion, we see that the optical axis deviates by the amount of $\Delta\vec r^\perp_\odot$ that is estimated from $\Delta\vec r^\perp_\odot/(z_0+z)= \vec r^\perp_\odot/z_0$. Adding to this deviation the orbital projection scaled in accord to the SGL imaging properties, given by (\ref{eq:S_z*6z-pos}), results in (\ref{eq:pl-pos-p2+perp-k})--(\ref{eq:hs-pos-p2-perp-k}).

\subsubsection{Mission-related assumptions}

Finally, we need some assumptions relevant to the SGL imaging mission, especially in regards to the velocity of the imaging telescope and the overall mission duration. For that, we assume that the imaging telescope does not stop at any given heliocentric position, but continues its egress from the solar system with a constant heliocentric velocity. In other words, we treat $z(t)=547\,{\rm AU}+v_{\tt sc} t$, where $v_{\tt sc}=25$ AU/yr. This assumption is needed to consider the temporal variability in the image position in the image plane at any given moment. As the telescope recedes from the Sun, it samples the project image in different image planes, positioned at different heliocentric distances. According to (\ref{eq:S_z*6z-pos}), this egress motion results in an increase of the image size. This fact must be accounted for in the mission design.  Although it is assumed that the imaging phase of the mission begins at 650 AU (when the solar coronagraph will begin reaching its optimal performance in rejecting the noise due to the brightness of the solar corona, see \cite{Turyshev-Toth:2020-extend,Toth-Turyshev:2020}) and continues for the duration of $10$~yr \cite{Turyshev-etal:2020-PhaseII}, in our analysis here we will use a longer duration of an extended science mission phase of $\Delta t=20$~yr. During this time the telescope, moving with $25$ AU/yr, will traverse the range of heliocentric distances between 550--1050 AU. Considering such an extended science mission allows us to model one full orbital period of Jupiter, which plays the dominant role in the reflex motion of the Sun and, given our  exoplanetary system assumptions, also of the host star, albeit with the different initial conditions.

Ultimately, for this mission we may have to consider effects of the second order, such as the reflex motion due to other celestial bodies in the solar system, mutual planetary interactions and also general relativistic effects \cite{Turyshev-Toth:2013}, as well as the actual dynamical properties of the target planetary system. These are beyond our current objective, which is to estimate establish upper limits on the navigational requirements for an imaging telescope, which can then be translated into estimates on the required onboard resources for a spacecraft carrying such a telescope to the SGL focal region and facilitating its operations. For this, the modeling of our solar system and the exoplanetary system as described above are sufficient. However, with establishment and maintenance of the local reference frame (will be discussed in Sec.~\ref{sec:RF-and-imaging}) these effects may be of less important and accounting only the leading terms may be sufficient.

\section{Image positions in the image plane}
\label{sec:position}

\subsection{Modeling position vectors}

To describe the displacement of images in the image plane, we need to consider positions of the primary optical axes given by (\ref{eq:pl-pos-p2+perp-k}) and (\ref{eq:hs-pos-p2-perp-k}). To aid with numerical analysis, in Appendix~\ref{sec:Kepler}, we present expressions needed to evaluate the orbits of the exoplanetary system, in the form of the position, velocity and acceleration vectors.  We use those expressions to evaluate the dominant dynamics on the image plane as relevant to a prospective imaging campaign at the SGL focal region.

First, we consider the motion of our solar system. The primary effect here results from the reflex motion of our Sun under the gravitational pull of the giant planets, known as the solar wobble.  For this, we use a simplified model for the solar reflex motion (i.e., accounting only for the largest effects) that does not include planet-planet interactions or relativistic corrections. Such a model may be given as
{}
\begin{eqnarray}
{\vec r}_\odot(t) = -
  \sum_j \frac{m_j}{M_j}r_j(t)+{\cal O}\big(c^{-2}\big),
 \label{eq:sol-ref-mot}
 \end{eqnarray}
where $M_j=m_\odot+m_j$, where $j=\{{\tt J,S, N,U}\}$ and $r_j(t)$ are the barycentric position vectors of the planets. We omitted the inner solar system planets in the sum, as their contributions to the reflex motion of the Sun are negligible at our level of approximation. We use (\ref{eq:pos-mXt})--(\ref{eq:pos-mZt}), to identify the orientation and temporal behavior of the planetary orbits, $r_j(t)$, and present the equation describing the reflex motion of the Sun with respect to the SSB frame as
{}
\begin{eqnarray}
{\vec r}_\odot(t)
 &=&
    \left( \begin{aligned}
x^0_\odot& \\
y^0_\odot& \\
z^0_\odot &\\
  \end{aligned} \right) -
  \sum_j \frac{m_j}{M_j}a_j \bigg\{
      \left( \begin{aligned}
A_j& \\
B_j& \\
C_j &\\
  \end{aligned} \right)
  \Big(\cos E_j(t)-e_j\Big)+
        \left( \begin{aligned}
F_j& \\
G_j& \\
H_j &\\
  \end{aligned} \right)\sqrt{1-e_j^2}\sin E_j(t)\bigg\},
  \label{eq:pos-mXt_s}
\end{eqnarray}
where $a_j$ and $e_j$ are the semi-major axis and  eccentricities of the planetary orbits involved, the Thiele--Innes constant orbital elements $A_j,B_j,...H_j$ are given by (\ref{eq:pos-A})--(\ref{eq:pos-H}) and the time  dependence is via the eccentric anomaly $E_j(t)$ given by (\ref{eq:pos-M3}). Also, ${\vec r}_\odot^0=(x^0_\odot, y^0_\odot, z^0_\odot)$ are the initial conditions.

Next, similarly to the approach used to discuss the solar reflex motion, we develop the expressions needed to describe the position vector of the host star with respect to its own barycentric coordinate reference frame. In addition, the host star position is changing due to the star's motion with respect to the Sun that includes components of the proper motion, $(\mu_\delta,\mu_\alpha)$, and radial velocity, $v_r$. The corresponding expressions for the host star postion vector have the form ${\vec r}_{\tt hs}(t)=(x_{\tt hs}(t), y_{\tt hs}(t), z_{\tt hs}(t))$, with the following components:
{}
\begin{eqnarray}
{\vec r}_{\tt hs}(t)
 &=&
z_0    \left( \begin{aligned}
\Delta \delta_0+\mu_\delta (t-t_0) & \\
\Delta \alpha_0+\mu_\alpha (t-t_0)& \\
1+v_r(t-t_0)/z_0 &\\
  \end{aligned} \right) -
  \sum_j \frac{m_j}{M_j}a_j \bigg\{
      \left( \begin{aligned}
A^{\tt exo}_j& \\
B^{\tt exo}_j& \\
C^{\tt exo}_j&\\
  \end{aligned} \right)
  \Big(\cos E_j(t)-e_j\Big)+
        \left( \begin{aligned}
F^{\tt exo}_j& \\
G^{\tt exo}_j& \\
H^{\tt exo}_j&\\
  \end{aligned} \right)\sqrt{1-e_j^2}\sin E_j(t)\bigg\},~~~~~
  \label{eq:pos-mXt_hs}
\end{eqnarray}
where  the Thiele-Innes constant orbital elements $A^{\tt exo}_j,B^{\tt exo}_j,...H^{\tt exo}_j$ are those for the solar system planets but scaled with the parameter $q=0.75\times10^6$ as $\Omega^{\tt exo}_j= q\, \Omega_j$, $\omega^{\tt exo}_j= q\, \omega_j$ and $i^{\tt exo}_j= q\, i_j$, yielding, in particular, $i^{\tt exo}_\oplus\simeq 37.5^\circ$.

Finally, we consider the contribution of the target exoplanet in the establishment of the primary optical axis toward the image of the exoplanet. To do that, we model the motion of the exoplanet as a part of the binary system -- host star and exo-Earth -- that does not interact with the other planets in the host star system. Again, using  (\ref{eq:pos-mXt}) and (\ref{eq:pos-mZt}), we have the exoplanet position vector, $\vec r_{\tt p}(t)=(x_{\tt p}(t),y_{\tt p}(t), z_{\tt p}(t))$, given as
{}
\begin{eqnarray}
\vec r_{\tt p}(t) &=&
      \left( \begin{aligned}
x^0_{\tt p}& \\
y^0_{\tt p}& \\
z^0_{\tt p} &\\
  \end{aligned} \right)
  +\frac{m_\odot a_{\tt E}}{m_\odot+m_{\tt E}} \bigg\{
      \left( \begin{aligned}
A^{\tt exo}_{\tt E} & \\
B^{\tt exo}_{\tt E} & \\
C^{\tt exo}_{\tt E} &\\
  \end{aligned} \right)
 \Big(\cos E_{\tt E}(t)-e_{\tt E}\Big)+
        \left( \begin{aligned}
F^{\tt exo}_{\tt E} & \\
G^{\tt exo}_{\tt E} & \\
H^{\tt exo}_{\tt E} &\\
  \end{aligned} \right) \sqrt{1-e_{\tt E}^2}\sin E_{\tt E}(t)\bigg\},~~~~~
  \label{eq:pos-mXt_p}
\end{eqnarray}
where all the orbital elements are for the binary host star system whose angles are uniformly amplified by the parameter $q$, according to $\{A^{\tt }_{\tt E}....H^{\tt }_{\tt E}\}\rightarrow \{A^{\tt exo}_{\tt E}....H^{\tt exo}_{\tt E}\}$,  as discussed in Sec.~\ref{sec:syst-mod}.
Also, ${\vec r}_{\tt p}^0=(x^0_{\tt p}, y^0_{\tt p}, z^0_{\tt p})$ are the initial conditions.

\subsection{The solar reflex motion}

\begin{figure}[h!]
\vskip 8pt
 \begin{center}
 \rotatebox{90}{\hskip 90pt  $\Delta \alpha$,  [nrad]}
\hskip -17pt
\begin{minipage}[b]{.42\linewidth}
\rotatebox{0}{\hskip -100pt  $z_0=10~${\rm ly}}
\vskip -16pt
 \includegraphics[width=0.810\linewidth]{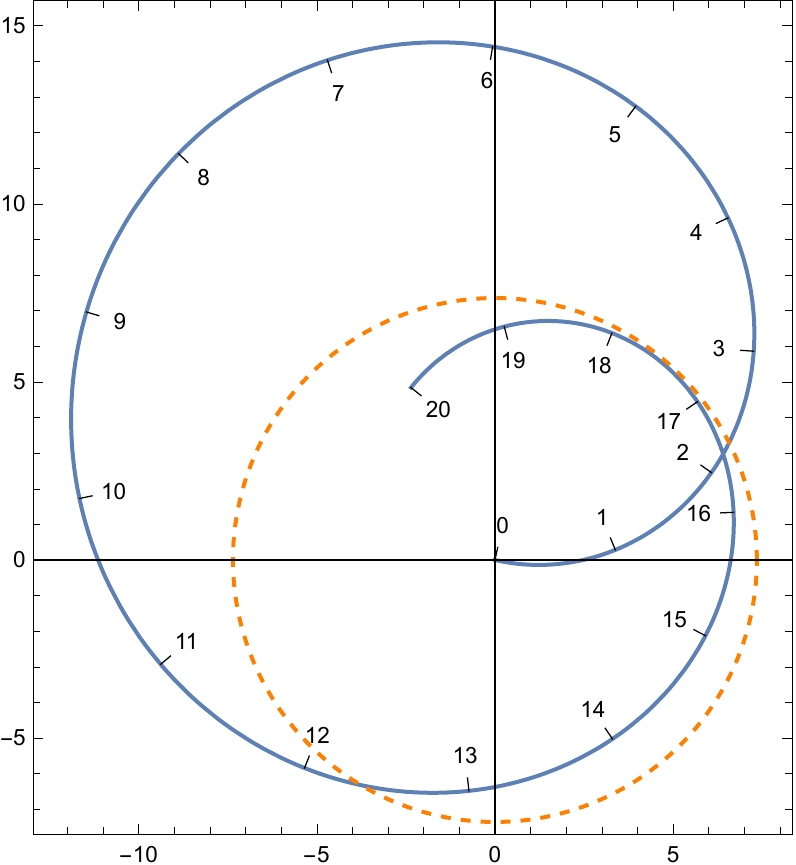}
 \rotatebox{0}{\hskip 30pt  $\Delta \delta$,  [nrad]}
\end{minipage}
  \hskip 12pt
   \rotatebox{90}{\hskip 60pt  Displacement, [$10^6$ km]}
\hskip -17pt
\begin{minipage}[b]{.42\linewidth}
\rotatebox{0}{\hskip -100pt  $z_0=10~${\rm ly}}
\vskip -16pt
 \includegraphics[width=0.83\linewidth]{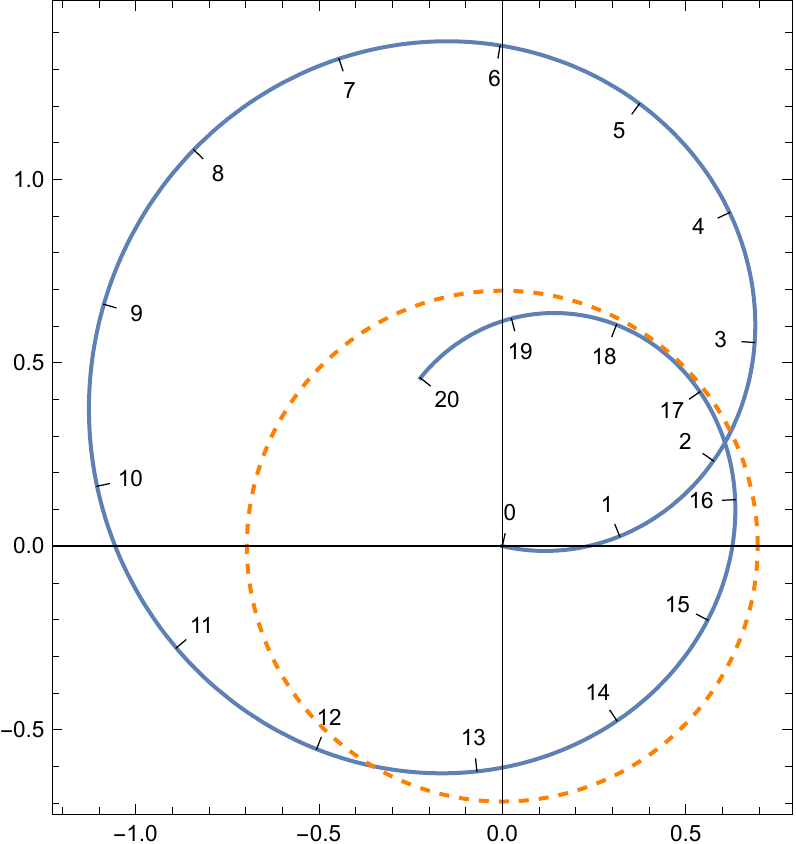}
 \rotatebox{0}{\hskip 40pt   Displacement, [$10^6$ km]}
\end{minipage}
\vskip -3pt
\caption{\label{fig:sun-reflex-astron} The impact of the solar wobble on the imaging geometry. Left: astrometric signal corresponding to the position of the solar optical center as given by (\ref{eq:sol-astrom_pos}) for $z_0=10$~ly. Ticks on the curve are in years. The dotted orange circle is the Sun's angular extent at that distance yielding $\delta\alpha^{\tt 10\,ly}_\odot=2R_\odot/z_0=14.7$~nrad. Right: physical displacement of the primary optical axis on the image plane corresponding to the astrometric signal on the left, as given by (\ref{eq:sol-r_pos}). The signal is projected onto the image plane that is moving at $v_{\tt sc}=25$ AU/yr. The dotted orange circle is the Sun's projected dimensions onto the image plane at $z=550$~AU, corresponding to $2r^{\tt 10\,ly}_\odot=(1+z/z_0) 2R_\odot \simeq 1.39\times 10^6$~km.
}
 \end{center}
\end{figure}

We begin the analysis by considering the impact of the solar reflex motion on the imaging geometry as given by (\ref{eq:pl-pos-p2+perp-k}) and (\ref{eq:hs-pos-p2-perp-k}).  For an observer at the distance $z_0=10$~ly, the angular displacement of the Sun due to the reflex motion of the Sun (i.e., solar wobble) under the gravitational pull from the giant planets is given as
{}
\begin{equation}
\Delta\vec \alpha_\odot(t)= \frac{\vec r^\perp_\odot}{z_0}.
\label{eq:sol-astrom_pos}
\end{equation}
Fig.~\ref{fig:sun-reflex-astron} (left) shows this signal for $z_0=10$~ly  as a function of time for the duration of $\Delta t=20$ years with the ticks indicating yearly marks. The dotted orange circle is the Sun's angular size at seen from $z_0$, which is estimated to be
$\delta\alpha_\odot=2R_\odot /z_0 =1.47\times 10^{-8} \,\big({10\,{\rm ly}}/{z_0}\big)~{\rm rad}.$
 Note the uneven spacing in the yearly tick marks. This effect is due to various planetary periods characterizing the solar system planets considered in this study (see Table~\ref{tab:ssplanets}).

Displacement of the Sun displaces the physical position of the primary optical axis on the image plane. As, after passing by the Sun, light continues toward the image plane at heliocentric distance of $z$. The resulted actual physical displacement of the axes is given by the first term in (\ref{eq:pl-pos-p2+perp-k}) and (\ref{eq:hs-pos-p2-perp-k}), namely
{}
\begin{equation}
\Delta\vec r^\perp_\odot(t)=  \Big(1+\frac{z(t)}{z_0}\Big)\vec r^\perp_\odot.
\label{eq:sol-r_pos}
\end{equation}

Fig.~\ref{fig:sun-reflex-astron}  (right) shows the magnitude of this displacement as a function of time. For that, we assumed a telescope egress velocity of $v_{\tt sc}=25$~AU/yr yielding $z(t)=547~{\rm AU}+v_{\tt sc}\cdot t+{\cal O}(t^2)$. This displacement is modeled as the deviation of the solar center's projection on the image plane from its initial position at 550~AU. The dotted orange circle in the plot indicates the Sun's projected size of
$\big(1+{z(t)}/{z_0} \big)2R_\odot \simeq 1.39  \times 10^6~{\rm km}.$ Again, the ticks are in years.

Clearly, the first term in (\ref{eq:sol-r_pos}) dominates the overall contribution from the solar reflex motion. This term is same from for any target. As the solar motion with respect to the SSB is rather well established, the relevant dynamics of the primary optical axis induced by this term may be addressed by the choice of an appropriate onboard propulsion system. Specific performance of such a system constitutes the coarse navigational requirement for the ultimate mission.

The next step is to evaluate the behavior of the solar reflex motion scaled with $z/z_0$ ratio, namely the $({z(t)}/{z_0})\vec r^\perp_\odot$ term, as shown in (\ref{eq:sol-r_pos}). The presence of the scaling ratio provides for an interesting temporal variability. There is an extra displacement of the optical axis that is due to changes in the image plane as a result of the ever increasing heliocentric distance.

Note that the plots in Fig.~\ref{fig:sun-reflex-astron} include contributions due to the increasing distance between the telescope and the Sun.  As a result, the physical size of the solar wobble due to $({z(t)}/{z_0})\vec r^\perp_\odot$ term projected on successive planes is increasing with heliocentric distance, causing the different curvatures of the curves shown on two plots. Clearly, the slower the egress motion, the less pronounced is its contribution to the change in the projected position. In any case, the effect of the reflex motion of the Sun on the position of the primary optical axes of the exoplanet and the host star remains dominant, with the telescope's egress motion providing only a minor contribution.

The presence of the second term in (\ref{eq:sol-r_pos}) provides the opportunity to appreciate and consider the smaller effects that our mission design will have to be able to address. If the magnitude of the largest term in (\ref{eq:sol-r_pos}) may be estimated based on Jupiter's gravitational pull on the Sun and evaluated to be  $r^\perp_\odot \simeq
m_J/(m_\odot+m_J)a_J\simeq 7.42  \times 10^5~{\rm km}$, the scaled term sets a smaller magnitude of the relevant effects $(z/z_0)r^\perp_\odot \simeq
7.63  \times 10^2~{\rm km}~(z/650~{\rm AU})(10~{\rm ly}/z_0) $. Below, we use the same logic to introduce and discuss other motions that are relevant to the SGL mission design.

\subsection{Reflex motion of the host star}

\begin{figure}[ht]
\vskip 8pt
 \begin{center}
 \rotatebox{90}{\hskip 80pt  $\Delta \alpha$,  [nrad]}
\hskip -17pt
\begin{minipage}[b]{.44\linewidth}
\rotatebox{0}{\hskip -95pt  $z_0=10~${\rm ly}}
\vskip -16pt
 \includegraphics[width=0.835\linewidth]{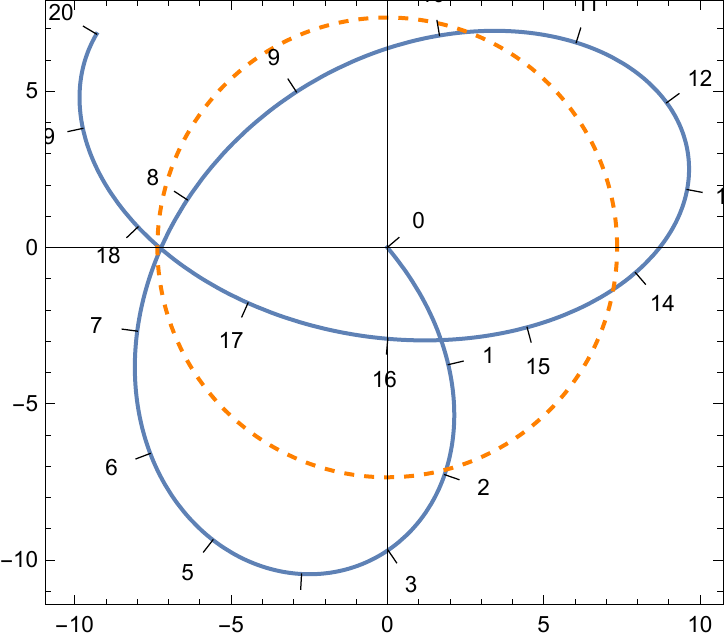}
 \rotatebox{0}{\hskip 20pt  $\Delta \delta$,  [nrad]}
\end{minipage}
  \hskip 7pt
   \rotatebox{90}{\hskip 50pt  Displacement, [$10^3$ km]}
\hskip -17pt
\begin{minipage}[b]{.5\linewidth}
\rotatebox{0}{\hskip -100pt  $z_0=10~${\rm ly}}
\vskip -16pt
 \includegraphics[width=0.84\linewidth]{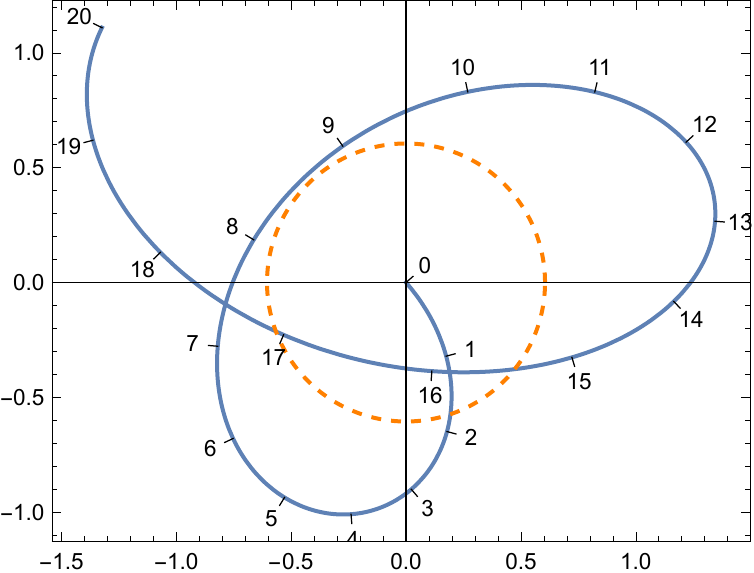}
 \rotatebox{0}{\hskip 30pt   Displacement, [$10^3$ km]}
\end{minipage}
\vskip -3pt
\caption{\label{fig:sun-reflex-astron-hs} Reflex motion (wobble) of the host star as seen from the distance of $z_0=10$~pc. Left: astrometric signal. Ticks on the curve are in years, corresponding to an egress velocity of $v=25$~AU/yr. The dotted orange circle angular size of the host star given as $\Delta\alpha_{\tt hs}= 2R_\odot/z_0=14.7$~nrad. The proper motion of the host star is assumed to be $\mu=\sqrt{\mu_\alpha^2+\mu^2_\delta}\lesssim 1~\mu$as/yr.
Right: the same reflex motion of the host star as in the left, but shown as the displacement from the optical axis, projected on the moving image plane. The dotted orange circle is the physical extent of the host star image as seen from $z=550$~AU, corresponding $r_\odot=(z/z_0) R_\odot $, resulting in a projected radius of $r^{\tt 10\,ly}_\odot=605.0$~km in the image plane.}
\vskip 16pt
 \rotatebox{90}{\hskip 15pt  Heliocentric distance,  [AU]}
\hskip -5pt
\begin{minipage}[b]{0.55\linewidth}
\rotatebox{0}{\hskip -90pt  $z_0=10~${\rm pc}}
\vskip -16pt
\includegraphics[width=0.95\linewidth]{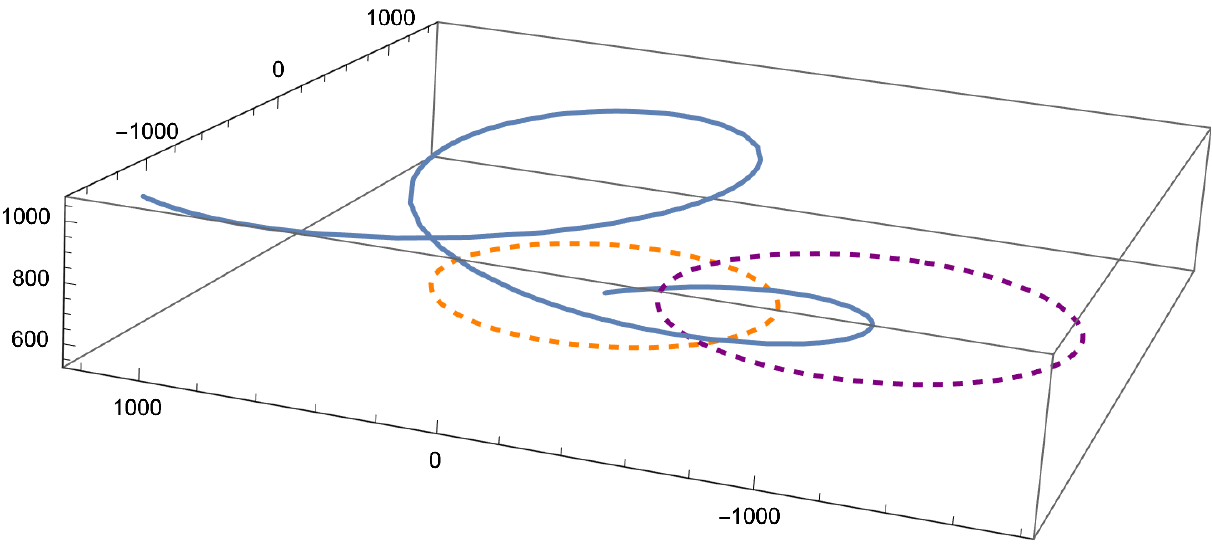}
 \rotatebox{0}{\hskip 10pt  Displacement, [km]}
\end{minipage}
\vskip -3pt
\caption{\label{fig:sun-reflex-3D-hs} The 3-D motion of the intersection of the host star's primary optical axis and the image plane defined by an egressing telescope, due to the reflex motion of the host star, as seen from $z_0=10$~ly. Dotted orange and purple circles represent the size of the host star, as projected onto the image plane at $z=550$~AU and 650~AU with $r_\odot=(z/z_0) R_\odot $ of $r^{\tt 10\,ly}_\odot=605.0$~km and $715.1$~km, correspondingly.
}
 \end{center}
\end{figure}

With the knowledge of the individual contributions of the Sun's reflex motion, that of the host star, and the orbital motion of the exoplanet, we may now consider their effects in defining the SGL optical axes toward the image of the host star and that of the target exoplanet.  We begin with the term in (\ref{eq:hs-pos-p2-perp-k}) representing the contribution of the host star:
{}
\begin{equation}
\Delta\vec r^\perp_{\tt hs}(t)=  -\frac{z(t)}{z_0}\vec r^\perp_{\tt hs}.
\label{eq:hs-r_pos}
\end{equation}

Fig.~\ref{fig:sun-reflex-astron-hs} (left) shows the astrometric signal due to the reflex motion (wobble) of the host star that is modeled as $\Delta\vec \alpha_{\tt hs}(t)=\Delta\vec r^\perp_{\tt hs}(t)/z_0=({z}(t)/{z_0})\vec r^\perp_{\tt hs}/z_0$, as seen from  $z_0=10$~ly. Ticks on the curve are in years, corresponding to an egress velocity of $v_{\tt sc}=25$~AU/yr. Dotted orange circle is the exo-Sun dimensions at $z=550$~AU, corresponding the angular size of $\Delta\alpha_{\tt hs}= 2R_\odot/z_0$, yielding $\Delta\alpha^{\tt 10\,ly}_{\tt hs}=14.7$~nrad. Here we assumed that the proper motion was compensated by the mission design to the level of $\mu=\sqrt{\mu_\alpha^2+\mu^2_\delta}\lesssim 1~\mu$as/yr. That is to say that the telescope trajectory was formed precisely to minimize the effect due to proper motion. As we show below, the effect of the proper motion on the image position is small and can be managed by the available onboard propulsion capabilities.

Fig.~\ref{fig:sun-reflex-astron-hs} (right) shows the same reflex motion, but presented as the physical displacement of the host star's primary optical axis, as projected onto the image plane while the telescope recedes from the solar system. The dotted orange circle is the image of the host star as seen from $z=550$~AU, corresponding to a circle with the radius of $r^{\tt 10\,ly}_\odot=605.0$~km. We note that the physical displacement of the optical axis due to the reflex of the host star is qualitatively different from that of the Sun. Clearly, the details of this motion are target specific, but should be well-known. Fig.~\ref{fig:sun-reflex-3D-hs} shows the 3-D motion of the host star's optical center due to reflex motion, corresponding to that shown on Fig.~\ref{fig:sun-reflex-astron-hs} (right).

Comparing the reflex motion of our Sun with that of the host star of the modeled exoplanetary system shown in Fig.~\ref{fig:sun-reflex-astron}
and Figs.~\ref{fig:sun-reflex-astron-hs}--\ref{fig:sun-reflex-3D-hs}, correspondingly, we notice drastically different behavior. This difference is due to the fact that the orbits of the giant planets in the model exoplanetary system are at very different inclinations not only to the solar system's ecliptic plane, but also with respect to each other. Therefore, they are pulling the host star in various directions (with the exo-Jupiter playing the dominant role) causing a distinct signature. This difference amplifies various dynamical effects, allowing us to explore the parameter space relevant for the mission design.

\begin{figure}[t!]
 \begin{center}
  \rotatebox{90}{\hskip 74pt  $\Delta \alpha$,  [nrad]}
\hskip -30pt
\begin{minipage}[b]{.40\linewidth}
\rotatebox{0}{\hskip -60pt  $z_0=10~${\rm ly}}
\vskip -16pt
 \includegraphics[width=0.72\linewidth]{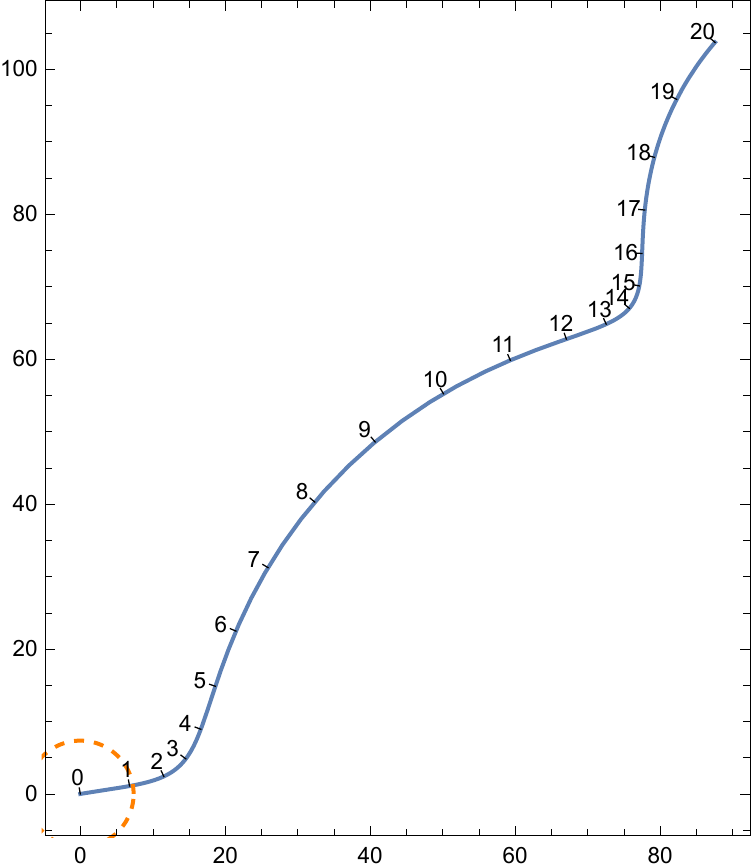}
 \rotatebox{0}{\hskip 27pt  $\Delta \delta$,  [nrad]}
\end{minipage}
  \hskip -5pt
   \rotatebox{90}{\hskip 50pt  Displacement, [$10^3$ km]}
\hskip -28pt
\begin{minipage}[b]{0.4\linewidth}
\rotatebox{0}{\hskip -50pt  $z_0=10~${\rm ly}}
\vskip -16pt
 \includegraphics[width=0.71\linewidth]{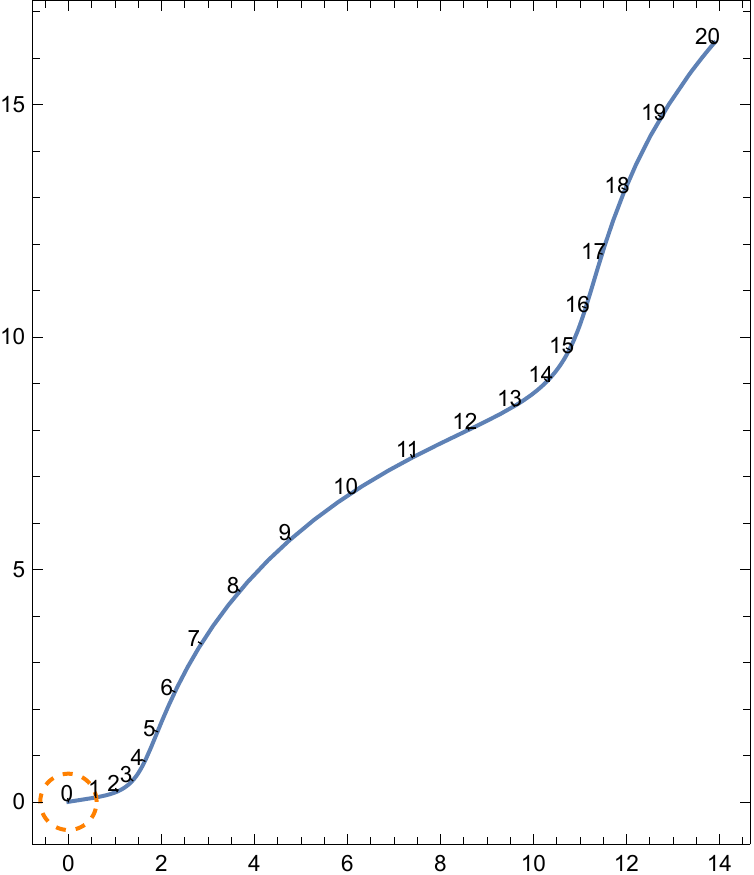}
 \rotatebox{0}{\hskip 20pt   Displacement, [$10^3$ km]}
\end{minipage}
\vskip -3pt
\caption{\label{fig:pm_1mas} Motion of the host star's SGL primary optical axis in the image plane (based on (\ref{eq:pos-mXt_hs})) as seen from $z_0=10$~ly and dominated by its proper motion at $\mu=1$~mas/yr. Left: the astrometric signal. Right: physical deviations from the initial position. The dotted orange circle is the host star's dimensions as seen from $z=550$~AU, similar to those in Fig.~\ref{fig:sun-reflex-astron-hs}.
}
\vskip 12pt
\hskip -17pt
 \rotatebox{90}{\hskip 83pt  $\Delta \alpha$,  [nrad]}
\hskip -17pt
\begin{minipage}[b]{.350\linewidth}
\rotatebox{0}{\hskip 95pt  $z_0=10~${\rm ly}}
\vskip -16pt
 \includegraphics[width=0.83\linewidth]{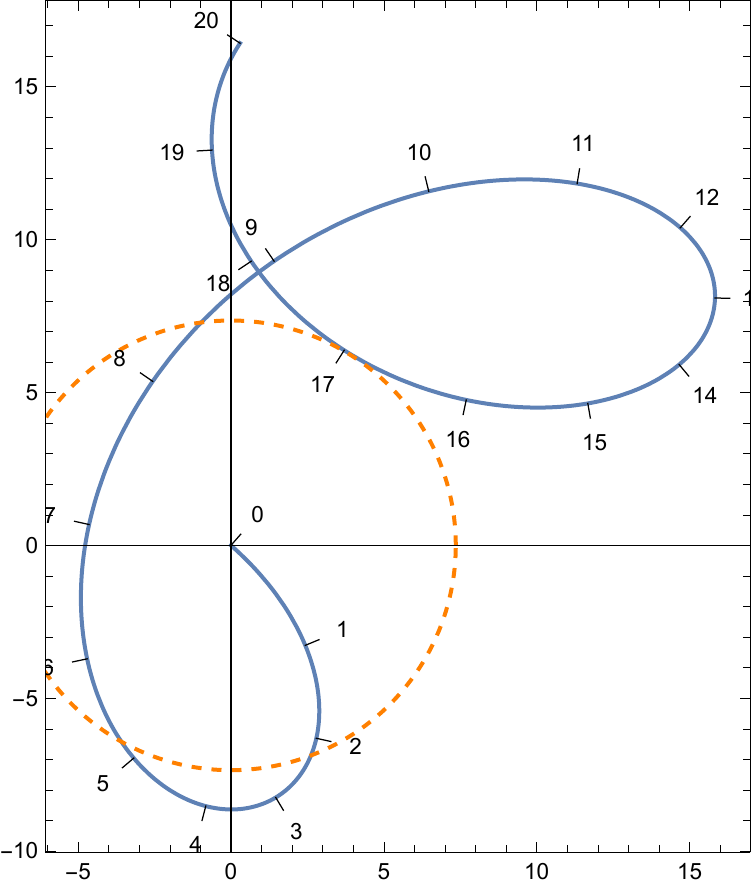}
 \rotatebox{0}{\hskip 26pt  $\Delta \delta$,  [nrad]}
\end{minipage}
  \hskip 15pt
   \rotatebox{90}{\hskip 55pt  Displacement, [$10^3$ km]}
\hskip -8pt
\begin{minipage}[b]{0.313\linewidth}
\rotatebox{0}{\hskip 87pt  $z_0=10~${\rm ly}}
\vskip -16pt
 \includegraphics[width=0.875\linewidth]{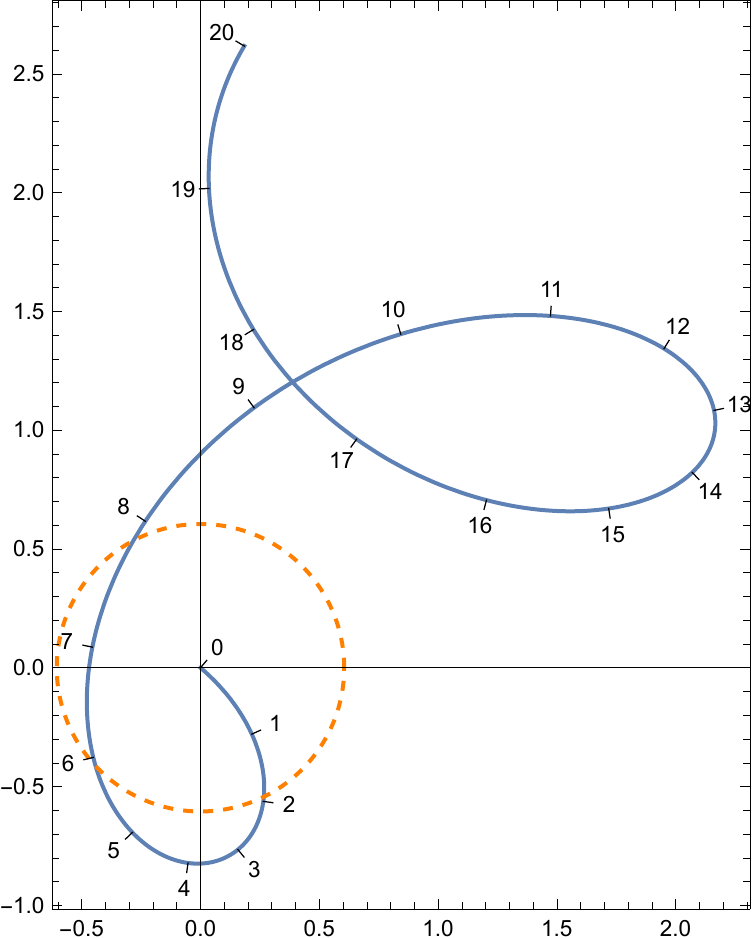}
 \rotatebox{0}{\hskip 25pt   Displacement, [$10^3$ km]}
\end{minipage}
\vskip -3pt
\caption{\label{fig:pm_100uas} Motion of the host star's SGL primary optical axis  in the image plane (as modeled by  (\ref{eq:pos-mXt_hs})), as seen from $z_0=10$~ly and dominated by its proper motion at $\mu=100~\mu$as/yr. Left: the astrometric signal. Right: deviations from the initial position. The dotted orange circle is the host star's dimensions as seen from $z=550$~AU, similar to those in Fig.~\ref{fig:sun-reflex-astron-hs}.
}
\vskip 8pt
 \rotatebox{90}{\hskip 0pt  Heliocentric distance, [AU]}
\hskip -12pt
\begin{minipage}[b]{.41\linewidth}
\rotatebox{0}{\hskip -70pt  $z_0=10~${\rm pc}}
\vskip -16pt
 \includegraphics[width=0.80\linewidth]{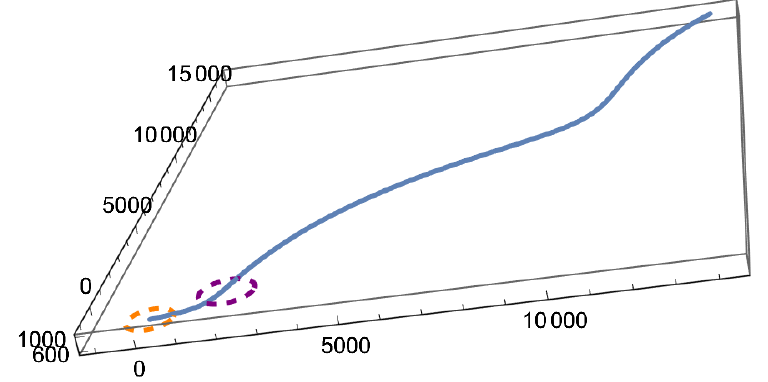}
 \rotatebox{0}{\hskip 20pt  Displacement, [km]}
\end{minipage}
  \hskip 12pt
   \rotatebox{90}{\hskip 0pt  Heliocentric distance, [AU]}
\hskip -12pt
\begin{minipage}[b]{.425\linewidth}
\rotatebox{0}{\hskip -100pt  $z_0=10~${\rm pc}}
\vskip -16pt
 \includegraphics[width=0.895\linewidth]{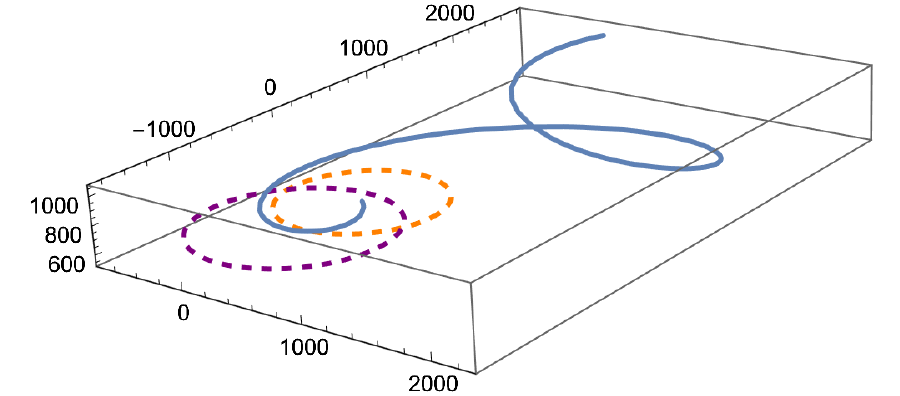}
 \rotatebox{0}{\hskip 20pt   Displacement, [km]}
\end{minipage}
\vskip -3pt
\caption{\label{fig:pm-3D} The 3-D motion of the host star's SGL primary optical axis on the image plane as modeled by  (\ref{eq:pos-mXt_hs}) and dominated by its proper motion at $\mu=1$~mas/yr (left, from Fig.~\ref{fig:pm_1mas}) and $\mu=100~\mu$as/yr (right, from Fig.~\ref{fig:pm_100uas}), as seen from $z_0=10$~ly.
}
 \end{center}
\end{figure}

We note that Figs.~\ref{fig:sun-reflex-astron-hs}--\ref{fig:sun-reflex-3D-hs} were produced under the assumption that the proper motion is rather well known and is  accounted for at the level of $\mu\leq 1~\mu$as/yr. This means that the telescope egresses in a precisely controlled direction that matches the proper motion of the host star. Rather than relying on this assumption, we also opted to relax it and instead study the effects of large values of uncompensated proper motion on the image position. Specifically, we allowed for the uncompensated proper motion to be as large as $\mu=1$~mas/yr and $\mu=100~\mu$as/yr.

Figs.~\ref{fig:pm_1mas}--\ref{fig:pm_100uas} show the motion of the host star's SGL primary optical axis under the influence of the proper motion, as modeled by (\ref{eq:pos-mXt_hs}). Left plots in both figures show the astrometric signal. The plots on the right present deviations from the initial position of the optical axis at 550~AU for the duration of 20 years.  As before, the ticks on the curves are in years, corresponding to an egress velocity of $v_{\tt sc}=25$~AU/yr. In addition, Fig.~\ref{fig:pm-3D} presents the 3-D motion of the host star as modeled by  (\ref{eq:pos-mXt_hs}) and dominated by the proper motion at $\mu=1$~mas/yr (Fig.~\ref{fig:pm_1mas}) and $\mu=100~\mu$as/yr (Fig.~\ref{fig:pm_100uas}), as seen from $z_0=10$~ly. We observe that the smaller the effect of the proper motion, the more pronounced is the dynamics of the exoplanetary system in the image position.

As we can see, in both cases considered above, the behavior of the exoplanetary system shown in Figs.~\ref{fig:pm_1mas}--\ref{fig:pm-3D} is completely overwhelmed by the proper motion. On the other hand, even such a large uncompensated proper motion produces a linear displacement of $\sim 10^3$~km per year, which is much less than the annual periodic displacement of $\sim 10^5$~km of the exoplanet in its orbit around the host star, as shown in Fig.~\ref{fig:exoE-10pc} (which, in turn, is dwarfed by the solar wobble).  Furthermore, fundamentally, this effect is an inertial motion (or with small secular variability \cite{Mignard:2003}), and also it is readily predicable.

Precise knowledge of the host star position within the image plane will play a critical role in establishing the local reference frame that is needed for imaging. The availability of such a  frame (see discussion in Sec.~\ref{sec:RF-and-imaging}), reduces the navigational sensitivity to the effects discussed in this subsection.

\subsection{Motion of the exo-Earth}

Ultimately, the primary objective of the SGL mission is to position the imaging telescope in the extended image of the target exoplanet and move within it in a pixel-by-pixel fashion \cite{Turyshev-Toth:2020-extend,Turyshev-etal:2020-PhaseII,Toth-Turyshev:2020}. For that, we need to consider the motion of the exoplanet with respect to the host star.  That motion may be described by using the orbital parameters of the exoplanet given by (\ref{eq:pos-mXt_p}). We use that expression to study the relevant orbital details.

Now we are ready to study behavior of the primary axis of the exoplanet. Accounting for the individual contributions due to the exoplanet and that of the host star, we consider their combined effect on the direction of the primary optical axis for the exoplanet. Once the local reference frame provided by the host star is established, our task is to find the exoplanet by tracking its orbit. Using the expressions (\ref{eq:pl-pos-p2+perp-k}) and (\ref{eq:hs-pos-p2-perp-k}), the relevant position vector is given as
{}
\begin{eqnarray}
\delta{\vec r}^\perp_{\tt p 0}={\vec r}^\perp_{\tt p 0}-{\vec r}^\perp_{\tt hs 0}&=&-
\frac{z}{z_0}\Big(\vec r^\perp_{\tt p}-\vec r^\perp_{\tt hs}\Big),
\label{eq:diff-pos-p2-perp-k}
\end{eqnarray}
with the individual positional vectors in the host star system's barycenter frame given by  (\ref{eq:pos-mXt_hs}) and (\ref{eq:pos-mXt_p}).

\begin{figure}[ht]
 \begin{center}
 \hskip 2pt
 \rotatebox{90}{\hskip 65pt  $\Delta \alpha$,  [$\mu$rad]}
\hskip -15pt
\begin{minipage}[b]{.32\linewidth}
\rotatebox{0}{\hskip 89pt  $z_0=10~${\rm ly}}
\vskip -17pt
 \includegraphics[width=0.85\linewidth]{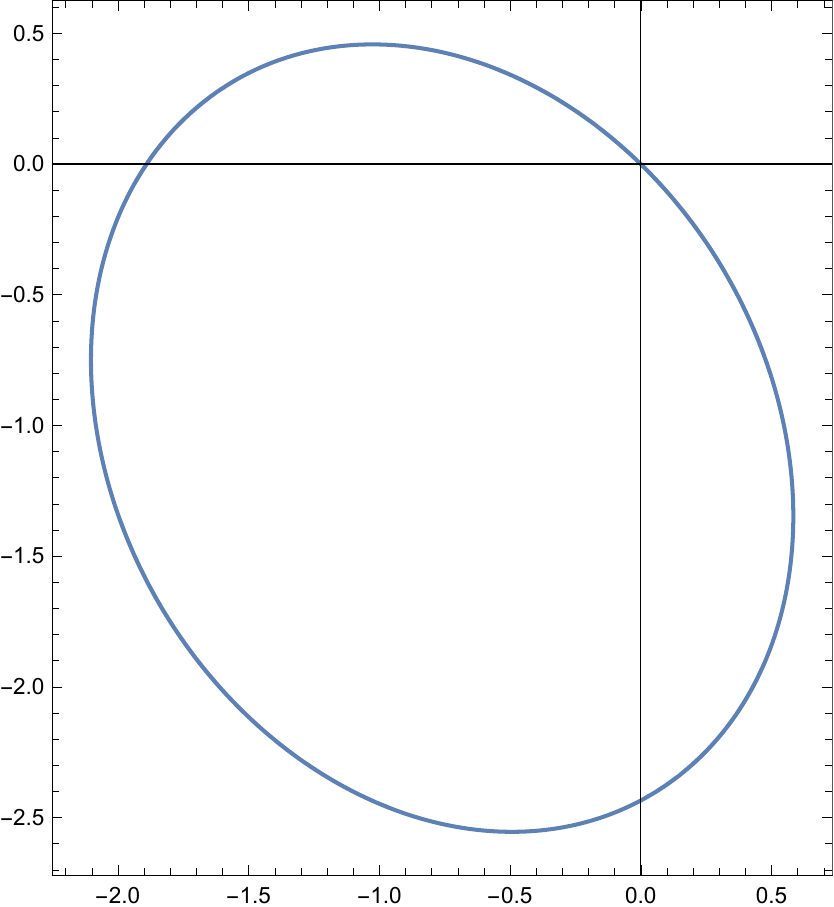}
 \rotatebox{0}{\hskip 22pt  $\Delta \delta$,  [$\mu$rad]}
\end{minipage}
  \hskip 5pt
   \rotatebox{90}{\hskip 40pt  Displacement, [$10^3$~km]}
\hskip -13pt
\begin{minipage}[b]{.32\linewidth}
\rotatebox{0}{\hskip 89pt  $z_0=10~${\rm ly}}
\vskip -17pt
 \includegraphics[width=0.86\linewidth]{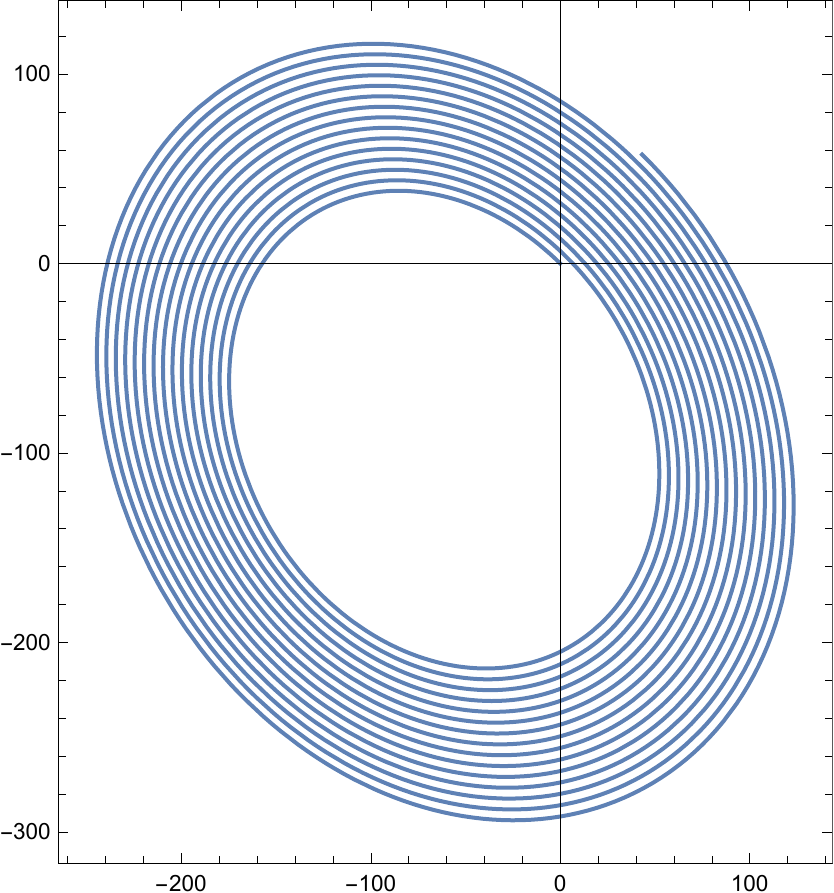}
 \rotatebox{0}{\hskip 10pt   Displacement, [$10^3$~km]}
\end{minipage}
  \hskip 5pt
   \rotatebox{90}{\hskip 40pt  Displacement, [$10^3$ km]}
\hskip -12pt
\begin{minipage}[b]{0.32\linewidth}
\rotatebox{0}{\hskip 89pt  $z_0=10~${\rm ly}}
\vskip -17pt
 \includegraphics[width=0.855\linewidth]{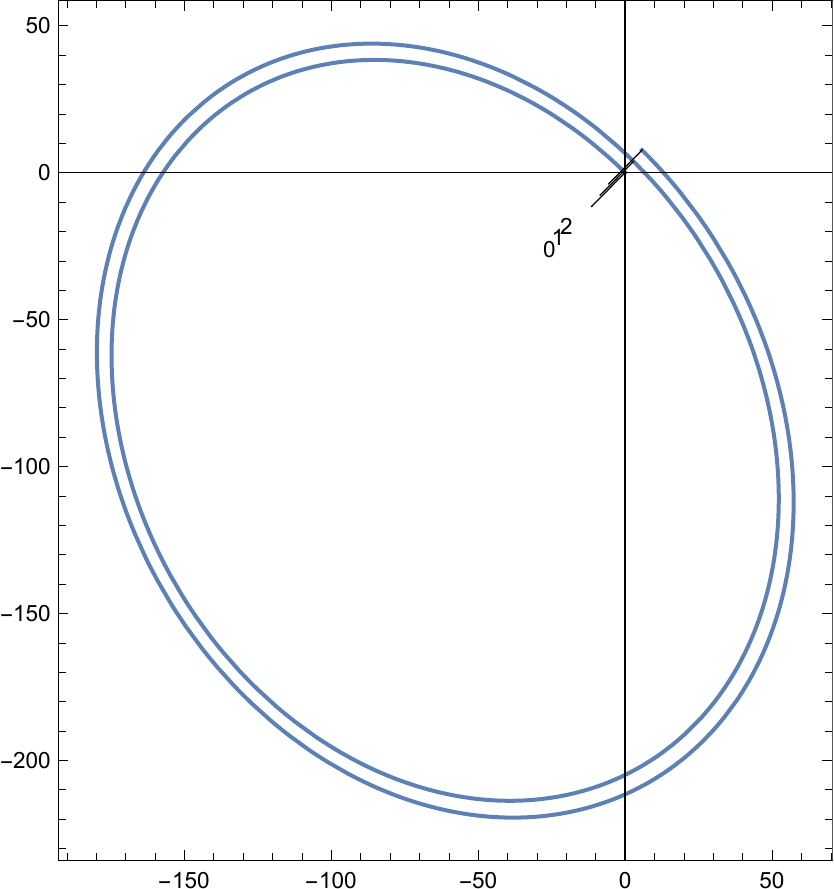}
 \rotatebox{0}{\hskip 30pt   Displacement, [$10^3$ km]}
\end{minipage}
\caption{\label{fig:exoE-10pc} Motion of the exo-Earth in its orbit around the host star at $z_0=10$~ly. Left: astrometric signal. Center: Projected image of the planet's orbit in the image plane defined by the egressing telescope.
Center: Projection of the exoplanetary orbit on an image plane that is moving at $v=25$~AU/yr, starting from $z=550$~AU for the duration  of 20 years. The orbit is referred to its initial position at the beginning of the observations. Right: the same as on the center plot, but shown only for 2 years.
}
\vskip 12pt
\hskip 2pt
    \rotatebox{90}{\hskip 30pt  Heliocentric distance,  [AU]}
     \hskip -25pt
\begin{minipage}[b]{.412\linewidth}
 \includegraphics[width=0.62\linewidth]{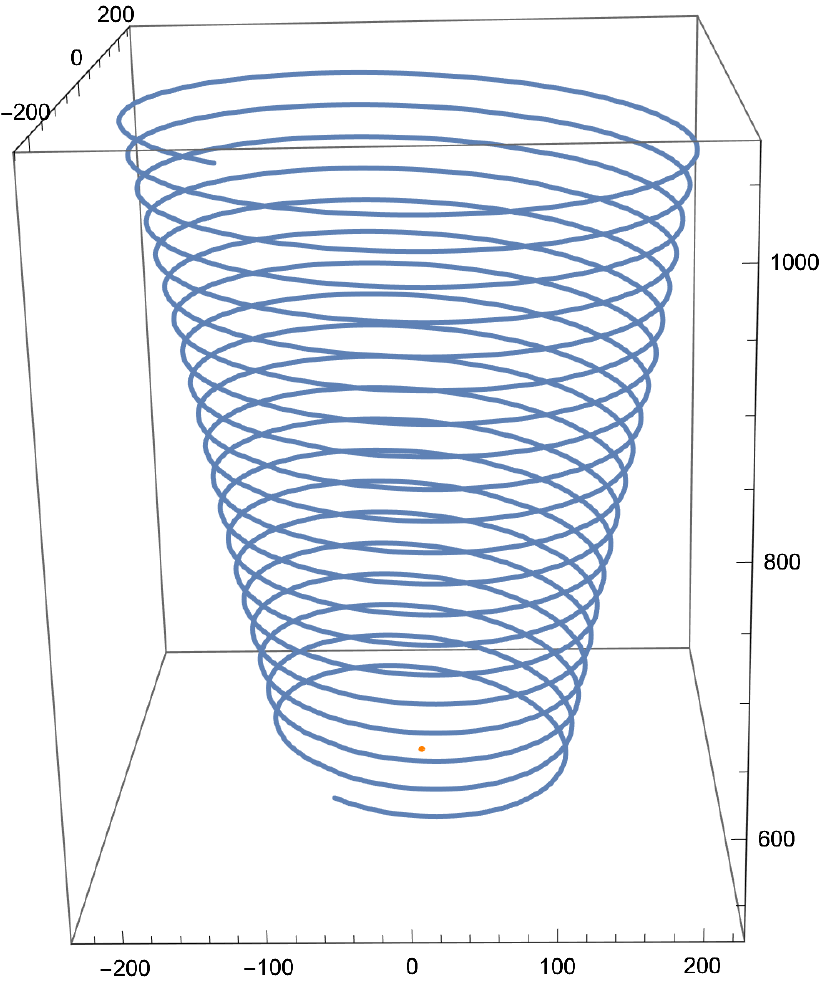}
 \rotatebox{0}{\hskip 5pt   Displacement, [$10^3$~km]}
\end{minipage}
\caption{\label{fig:exoE-3D-10ly} The 3D orbit of the exoplanet, relative to the host star and shown from different heliocentric distances.
}
\vskip 16pt
 \end{center}
\end{figure}

Fig.~\ref{fig:exoE-10pc} shows the motion of the exo-Earth in its orbit around the host star, as seen from $z_0=10$~ly.
In the left plot, we present the astrometric signal, at the center plot, we show physical deviations from the initial position (measured in $10^3$ km) of the projected location of the exoplanet's image in the image plane. As the size of the exoplanetary orbit projected on a moving image plane increases, the trajectory of the primary optical axis never closes and never repeats. In fact, it represents a tight spiral that opens up as mission progresses. The plot on the right zooms in on the same image as in the central panel, but for the initial 2 years of observations. It shows the planetary orbit relative to the host star. The center of the plot is the host star's position. Fig.~\ref{fig:exoE-3D-10ly} shows the 3-D position of the exoplanet's primary optical axis on the image plane, as seen from different heliocentric distances. Notice how the projected orbital dimensions increase as the heliocentric distance increases at the rate of the telescope motion at $v_{\tt sc}=25$ AU/yr.

We observe that both plots in Fig.~\ref{fig:exoE-10pc} (right)  show the same behavior but at different heliocentric ranges. Clearly, the astrometric signal is dominated by the ratio of $r_{\tt p}^\perp/z_0\simeq 1.58 ~\mu{\rm rad} \,(r_{\tt p}^\perp/1\,{\rm AU}) (10\,{\rm ly}/z_0)$, which stays the same thought the mission. However, according to (\ref{eq:diff-pos-p2-perp-k}), the size of the physical projection of the orbit on the image plane increases at the rate of $z(t)/z_0$, which is evident in the right plots on Fig.~\ref{fig:exoE-10pc}  and also quite pronounced in Fig.~\ref{fig:exoE-3D-10ly}.  Clearly, given the magnitudes of the two vectors, $\vec r^\perp_{\tt p}$ and $\vec r^\perp_{\tt hs}$, the exoplanetary orbit dominates in the expression (\ref{eq:diff-pos-p2-perp-k}).  Although the variability in the image position is quite large, being nearly 200 times larger than the host star's image size, it is predictable and moves in a steady manner.

We note that the exoplanet and therefore, its projection in the image plane are in continuous noninertial motion. In the next section, we evaluate the velocities and accelerations that characterize this motion.

\section{Image velocities and accelerations}
\label{sec:im-vel-acc}

In the previous section, we studied the various position vectors that determine behavior of the primary optical axes of the host star and that of the target exoplanet. As we saw, during the mission, these vectors typically will exhibit significant variability by deviating from their original positions by million of kilometers (i.e.,  solar reflex motion),  hundreds of thousands of kilometers (i.e., exoplanet) or even by tens of thousands of kilometers (i.e., host star). Such variability, if it occurred on short time scales, would represent a significant challenge for an SGL mission, resulting in realistically large velocity change requirements for an observing telescope.

As it turns out, while positional changes are large on annual timescales, instantaneous velocities and accelerations remain small. Nonetheless, precise knowledge of the velocities and accelerations of the optical axis are crucial for an imaging mission.

To understand why, consider the consequences of a navigational error. An error in position space amounts to a displacement of the image. So long as this displacement is modest, the implications are manageable: instead of being centered in the area of the image plane sampled by the imaging telescope, the projected image may be offset by some amount. Unless this offset is large enough so that substantial portions of the projected exoplanet image are missed altogether, the impact on image recovery of a positional error will be modest.

An error in velocity space has different consequences. To collect enough light to achieve the required signal-to-noise ratio, the imaging telescope may need to remain at a specific location, or pixel, for periods of time measured in minutes. If the telescope's velocity in the image plane does not match the velocity of the image, the result will be smearing and blur, substantially reducing the quality of the data collected and the chances of reconstructing a high quality image. Furthermore, as the telescope moves from location to location, from pixel to pixel, an unknown error in velocity would imply assigning the data entirely to the wrong pixel. This unintentional scrambling of the data set may render the image unrecoverable.

A precise parameterized model of the velocities and accelerations of the projected exoplanet image (with the parameters perhaps fitted and adjusted using the very observations made by these telescopes) in the image plane is therefore essential for good quality image reconstruction.

\subsection{Evaluating velocities}

We begin our discussion with the image velocities. To evaluate the contributions of the velocities of various objects involved in the imaging with the SGL, namely  $\dot{\vec r}_{\tt hs}, \dot{\vec r}_{\tt p}$ and $\dot{\vec r}_\odot$, we need to evaluate their behavior in the image plane. For that, we take the time derivative from (\ref{eq:pl-pos-p2+perp-k}) and (\ref{eq:hs-pos-p2-perp-k}), which yields
 the following expressions for the velocity of the primary optical axes in the image plane
 {}
\begin{eqnarray}
\dot {\vec r}^\perp_{\tt p 0}&=&\Big(1+\frac{z}{z_0}\Big)\dot {\vec r}^\perp_\odot+ \frac{v_{\tt sc}}{z_0}\vec r^\perp_\odot-\Big(
\frac{z}{z_0}\dot{\vec r}^\perp_{\tt p}+
\frac{v_{\tt sc}}{z_0}{\vec r}^\perp_{\tt p}\Big),
\label{eq:pl-perp-k-v}\\
\dot{\vec r}^\perp_{\tt hs 0}&=&\Big(1+\frac{z}{z_0}\Big)\dot{\vec r}^\perp_\odot+\frac{v_{\tt sc}}{z_0}{\vec r}^\perp_\odot-\Big(\frac{z}{z_0}\dot{\vec r}^\perp_{\tt hs}+\frac{v_{\tt sc}}{z_0}{\vec r}^\perp_{\tt hs}\Big),
\label{eq:hs-perp-k-v}
\end{eqnarray}
where, in order to evaluate the contribution of the galacto-centric motion, we assumed the line-of-sight velocity of $v_z=\dot z_0=200$~km/s, which for $z_0\simeq 10$~ly from (\ref{eq:pl-pos-p2+perp-k}) results in the induced image velocity on the order of $(v_z/z_0^2)zr_{\tt p}^\perp \sim(v_z/z_0^2) (650\,{\rm AU})(1\,{\rm AU})\lesssim 3.3\times 10^{-4}$~m/s, which is small for the purposes of our present objectives. Thus, we omit the $\dot z_0$-dependent term from the discussion.

\begin{figure}[ht!]
\vskip 5pt
 \begin{center}
 \rotatebox{90}{\hskip 65pt  Velocity, \,[m/s]}
\hskip -12pt
\begin{minipage}[b]{.39\linewidth}
\rotatebox{0}{\hskip -100pt  $z_0=10~${\rm ly}}
\vskip -16pt
 \includegraphics[width=0.85\linewidth]{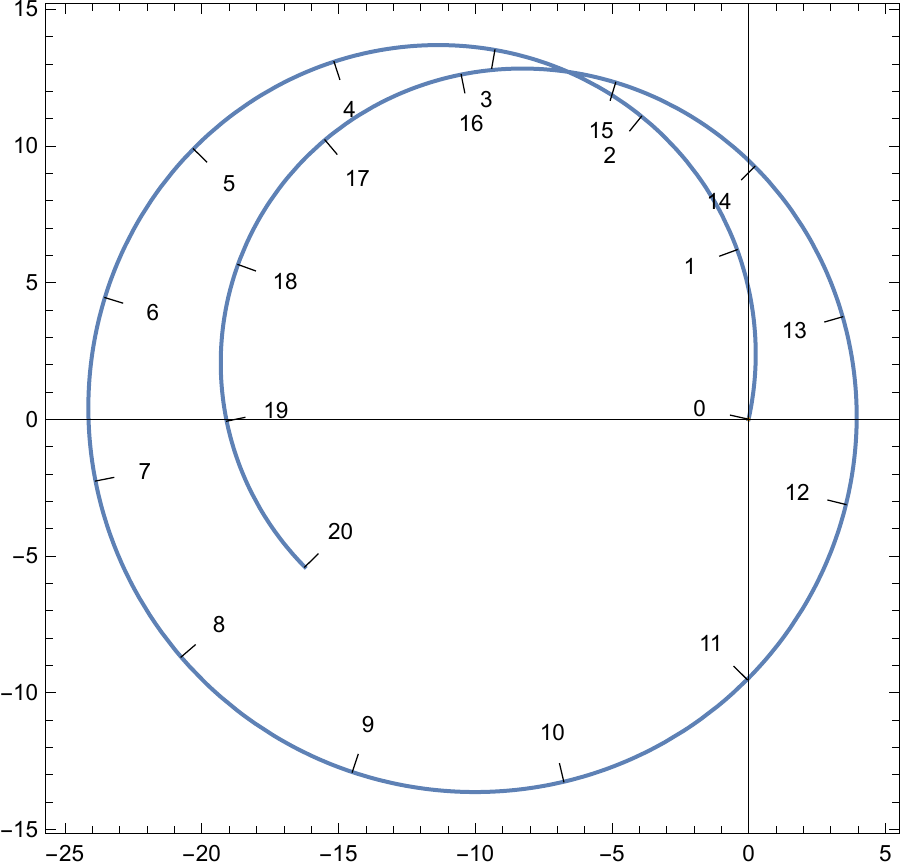}
 \rotatebox{0}{\hskip 25pt  Velocity, \,[m/s]}
\end{minipage}
  \hskip 10pt
   \rotatebox{90}{\hskip 65pt  Velocity, \,[m/s]}
\hskip -12pt
\begin{minipage}[b]{0.338\linewidth}
\rotatebox{0}{\hskip 95pt  $z_0=10~${\rm ly}}
\vskip -16pt
 \includegraphics[width=0.85\linewidth]{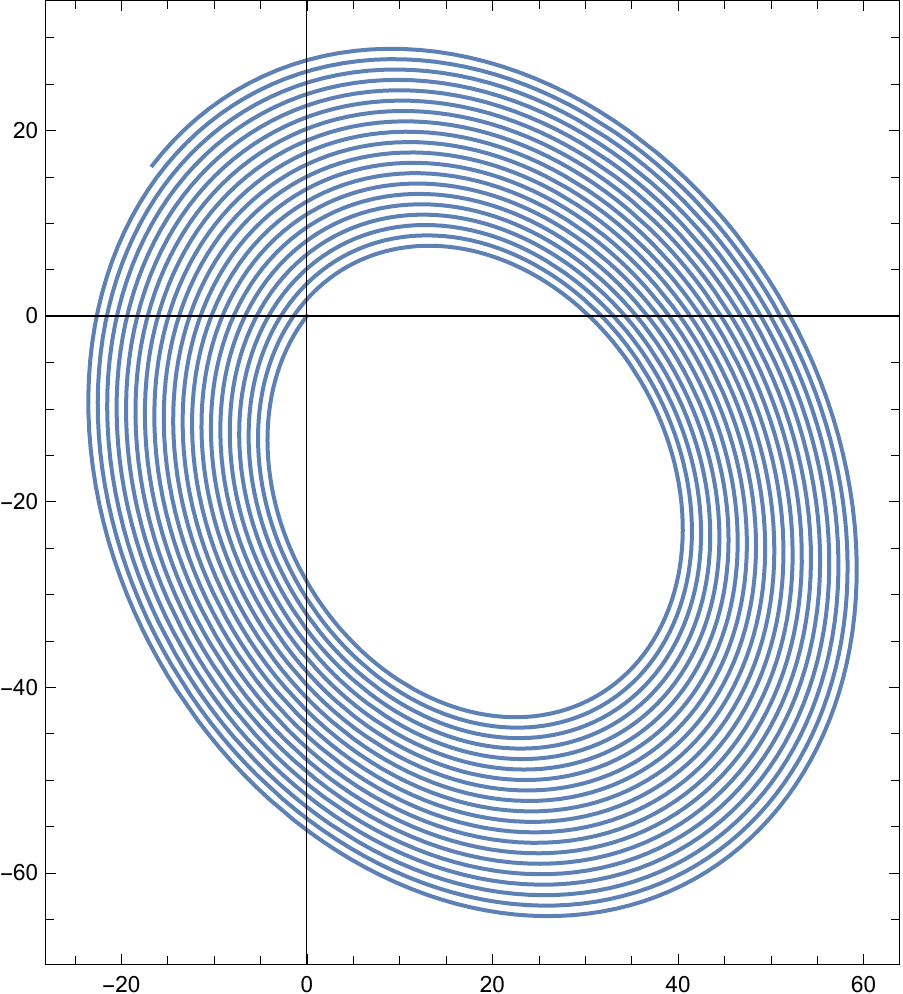}
 \rotatebox{0}{\hskip 25pt   Velocity, \,[m/s]}
\end{minipage}
\vskip 0pt
\caption{\label{fig:vel-SunExoE-10ly}  Left: velocity of the optical axis (referenced to initial value) due to reflex motions of the Sun for a target at $z_0=10$~ly. Ticks on the curve are in years, assuming that the image plane is moving at the rate of $v_{\tt sc}=25$ AU/yr. Right:  velocity of the optical axis (referenced to initial value) due to exo-Earth orbital motion, also for a target at $z_0=10$~ly, and projected onto the moving image plane. The inner ellipse corresponds to year 1 at $z=550$ AU, increasing outwards for 20 years.
}
 \end{center}
\vskip -2pt
 \begin{center}
 \rotatebox{90}{\hskip 60pt  Velocity, \,[m/s]}
\hskip -20pt
\begin{minipage}[b]{.40\linewidth}
\rotatebox{0}{\hskip 95pt  $z_0=10~${\rm ly}}
\vskip -16pt
 \includegraphics[width=0.75\linewidth]{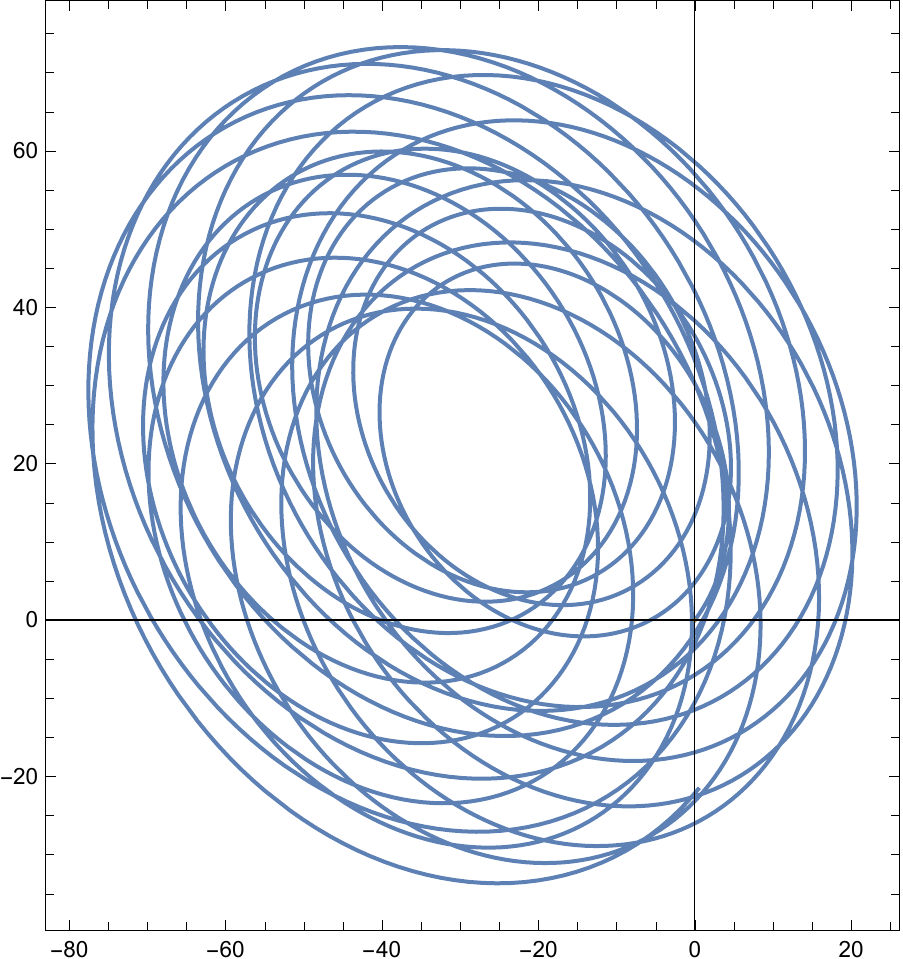}
  \rotatebox{0}{\hskip 20pt  Velocity, \,[m/s]}
\end{minipage}
  \hskip 10pt
\rotatebox{90}{\hskip 30pt  Heliocentric distance,  [AU]}
\hskip -50pt
\begin{minipage}[b]{0.33\linewidth}
\vskip -18pt
\includegraphics[width=0.335\linewidth]{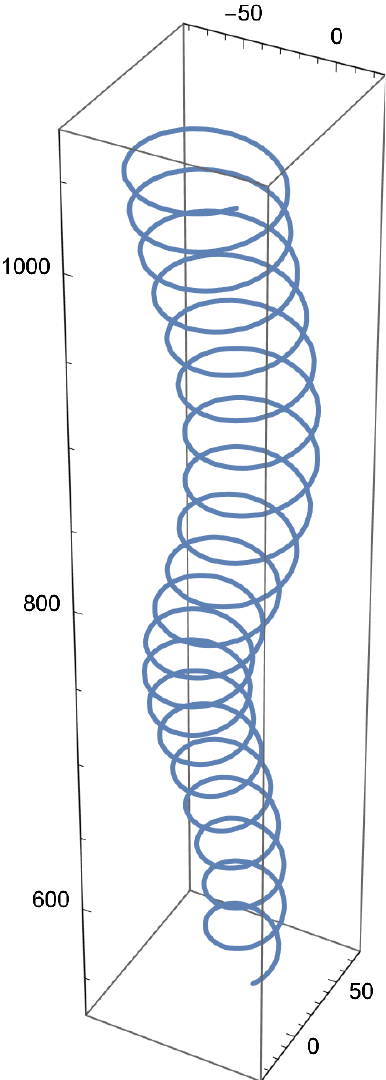}
\vskip 1pt
  \rotatebox{0}{\hskip 20pt  Velocity, \,[m/s]}
 \vskip 100pt
\end{minipage}
\vskip -3pt
\caption{\label{fig:vel-SunExoE-10ly3D}  Left: combined velocity of the optical axis due to reflex motions of the Sun and the exoplanet, as modeled by (\ref{eq:pl-perp-k-v}) with individual components shown in Fig.~\ref{fig:vel-SunExoE-10ly}.  Right:  same as on the left, but in a 3-dimensional projection for 20 years.
}
 \end{center}
\vskip 5pt
 \begin{center}
   \rotatebox{90}{\hskip 50pt  Velocity, \,[m/s]}
\hskip -10pt
\begin{minipage}[b]{0.45\linewidth}
\rotatebox{0}{\hskip 150pt  $z_0=10~${\rm ly}}
\vskip -16pt
 \includegraphics[width=0.86\linewidth]{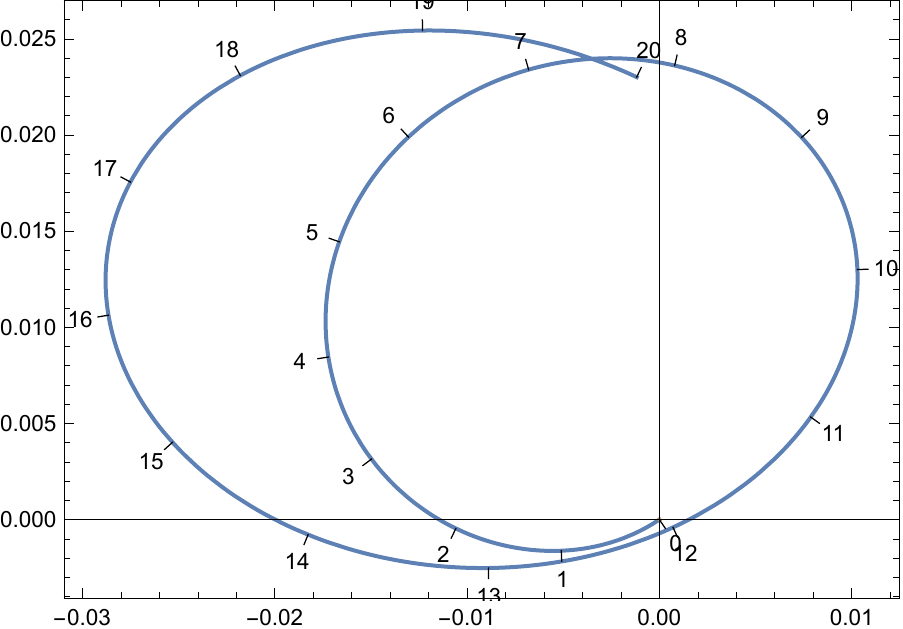}
 \rotatebox{0}{\hskip 30pt   Velocity, \,[m/s]}
\end{minipage}
\vskip 0pt
\caption{\label{fig:vel-HS-10ly} Induced velocity of the optical axis (referenced to initial value) due to the reflex motions of the host star, for a target at $z_0=10$~ly, and projected onto the image plane moving at $v_{\tt sc}=25$ AU/yr.}
 \end{center}
 \end{figure}

Using  ${\vec r}_\odot(t)$ from (\ref{eq:pos-mXt_s}) and  the results summarized by  (\ref{eq:pos-Xt-dot}), we develop the expression for the image velocity $\dot {\vec r}_\odot(t) $ as a result of the reflex motion of the Sun with respect to the SSB:
{}
\begin{eqnarray}
\dot {\vec r}_\odot(t) &=&
 \sum_j\frac{m_j}{M_j}\frac{n_ja_j}{1-e_j\cos E_j(t)} \bigg\{
 \left( \begin{aligned}
A_j& \\
B_j& \\
C_j &\\
  \end{aligned} \right)
\sin E_j(t) -
        \left( \begin{aligned}
F_j& \\
G_j & \\
H_j &\\
  \end{aligned} \right) \sqrt{1-e_j^2}\cos E_j(t)\bigg\}.~~~~~
  \label{eq:pos-mXt_sv}
\end{eqnarray}

Similarly, using the expression for ${\vec r}_{\tt hs}(t)$  from (\ref{eq:pos-mXt_hs}) and the results from (\ref{eq:pos-Xt-dot}), we have similar expressions  needed to evaluate the image velocities due to reflex motion of the host star:
{}
\begin{eqnarray}
\dot {\vec r}_{\tt hs}(t) &=&    \left( \begin{aligned}
z_0\mu_\delta& \\
z_0\mu_\alpha& \\
 v_r&\\
  \end{aligned} \right) +
 \sum_j\frac{m_j}{M_j}\frac{n_ja_j}{1-e_j\cos E_j(t)} \bigg\{
      \left( \begin{aligned}
A^{\tt exo}_j & \\
B^{\tt exo}_j & \\
C^{\tt exo}_j &\\
  \end{aligned} \right)
\sin E_j(t) -
        \left( \begin{aligned}
F^{\tt exo}_j & \\
G^{\tt exo}_j & \\
H^{\tt exo}_j&\\
  \end{aligned} \right) \sqrt{1-e_j^2}\cos E_j(t)\bigg\}.~~~~~
  \label{eq:pos-vx}
\end{eqnarray}

Finally, using ${\vec r}_{\tt p}(t) $ from (\ref{eq:pos-mXt_p}) and also (\ref{eq:pos-mXt_hs}), the velocity of the exo-Earth is modeled as
{}
\begin{eqnarray}
\dot {\vec r}_{\tt p}(t)
 &=&
-\frac{m_\odot}{m_\odot+m_{\tt E}}\frac{n_{\tt E}a_{\tt E}}{1-e_{\tt E}\cos E_{\tt E}(t)}\bigg\{
      \left( \begin{aligned}
A^{\tt exo}_{\tt E} & \\
B^{\tt exo}_{\tt E} & \\
C^{\tt exo}_{\tt E} &\\
  \end{aligned} \right)
\sin E_{\tt E}(t)
- \left( \begin{aligned}
F^{\tt exo}_{\tt E}& \\
G^{\tt exo}_{\tt E}& \\
H^{\tt exo}_{\tt E} &\\
  \end{aligned} \right) \sqrt{1-e_j^2}\cos E_{\tt E}(t)\bigg\}.~~~~~
  \label{eq:pos-vxp}
\end{eqnarray}

Fig.~\ref{fig:vel-SunExoE-10ly} shows the relevant velocities of the primary optical axes due to reflex motion of the Sun (left), $\Delta \dot {\vec r}^\perp_\odot$, and orbital motion of the exoplanet (right), $\Delta\dot {\vec r}^\perp_{\tt p}$, that from (\ref{eq:pl-perp-k-v}) have the from:
 {}
\begin{eqnarray}
\Delta \dot {\vec r}^\perp_\odot&=&\Big(1+\frac{z}{z_0}\Big)\dot {\vec r}^\perp_\odot+ \frac{v_{\tt sc}}{z_0}\vec r^\perp_\odot,
\qquad
\Delta\dot {\vec r}^\perp_{\tt p}=-\Big(
\frac{z}{z_0}\dot{\vec r}^\perp_{\tt p}+
\frac{v_{\tt sc}}{z_0}{\vec r}^\perp_{\tt p}\Big).
\label{eq:sun-pl-perp-k-v}
\end{eqnarray}
Specifically, the left plot shows the induced velocity of the optical axis (referenced to its initial value) due to the reflex motion of the Sun, $\Delta \dot {\vec r}^\perp_\odot$. Ticks on the curve are in years, assuming that the image plane moves with velocity of $v_{\tt sc}=25$ AU/yr.  The right plot presents the projected velocity of the optical axis (referenced to its initial value) due to exo-Earth's orbital motion (also, at $z_0=10$~ly), $\Delta\dot {\vec r}^\perp_{\tt p}$ and projected on the image plane moving with velocity of $v_{\tt sc}=25$ AU/yr. The inner ellipse corresponds to year 1 at $z=550$ AU, increasing outwards for 20 years. Clearly, the first terms in both expressions (\ref{eq:sun-pl-perp-k-v}) dominate, namely the $\dot{\vec r}^\perp_\odot$ and $({z}/{z_0})\dot{\vec r}^\perp_{\tt p}$, thus driving the mission requirements.

To appreciate the dynamics in the velocity space, Fig.~\ref{fig:vel-SunExoE-10ly3D}  shows the combined velocity of the optical axis (referenced to initial value) due to reflex motions of the Sun and the exoplanet, as modeled by (\ref{eq:pl-perp-k-v}) with individual components shown in Fig.~\ref{fig:vel-SunExoE-10ly}.  On the left, the 2D-view and on the right is the same, but in 3-dimensional projection for 20 years. This is a complex figure if the knowledge of the individual components is not available. However, this is not the case for the SGL, as all these motions will be well understood after a specific imaging target is selected (see discussion in Sec.~\ref{sec:a-priory}). Knowledge of the exoplanet's motion, as represented in this figure, will be essential to the success of any attempt to reconstruct an exoplanet image from multiple telescopic observations in the image plane.

Fig.~\ref{fig:vel-HS-10ly} shows the velocity of the primary optical axes induced by the reflex motion of the host star, $\Delta \dot {\vec r}^\perp_{\tt hs}$, that from (\ref{eq:hs-perp-k-v}) has the same form as  $\Delta\dot {\vec r}^\perp_{\tt p}$ given by (\ref{eq:sun-pl-perp-k-v}) where ${\vec r}^\perp_{\tt p}$ is replaced with ${\vec r}^\perp_{\tt hs}$. Note that the resulted velocity is smaller by a factor of $\sim10^3$, and is very small to contribute to the mission requirements.

Clearly, the motion of the Sun, $\dot {\vec r}^\perp_\odot$,  and the scaled velocity of exoplanet, $({z}/{z_0})\dot{\vec r}^\perp_{\tt p}$, induce velocity of the primary axis at the level of a few m/s requiring more attention from the mission design stand point. However, these velocities should not present challenges for modern onboard propulsion systems \cite{Turyshev-etal:2020-PhaseII}. The mission planning task will be to identify all these velocities in advance and compensate for their presence with appropriately chosen propulsion technology.

\subsection{Evaluating accelerations}

To evaluate the contributions of the accelerations $\ddot{\vec r}_{\tt hs}, \ddot{\vec r}_{\tt p}$ and $\ddot{\vec r}_\odot$ to the imaging, we need to study their behavior in the image plane. For that, we take the second time derivative from (\ref{eq:pl-pos-p2+perp-k}) and (\ref{eq:hs-pos-p2-perp-k}), which yields the following expressions for the accelerations of the primary optical axes in the image plane:
 {}
\begin{eqnarray}
\ddot {\vec r}^\perp_{\tt p 0}&=&\Big(1+\frac{z}{z_0}\Big)\ddot {\vec r}^\perp_\odot+2 \frac{v_{\tt sc}}{z_0}\dot{\vec r}^\perp_\odot-\Big(
\frac{z}{z_0}\ddot{\vec r}^\perp_{\tt p}+2
\frac{v_{\tt sc}}{z_0}\dot{\vec r}^\perp_{\tt p}\Big),
\label{eq:pl-perp-k-a}\\
\ddot{\vec r}^\perp_{\tt hs 0}&=&\Big(1+\frac{z}{z_0}\Big)\ddot{\vec r}^\perp_\odot+2\frac{v_{\tt sc}}{z_0}\dot{\vec r}^\perp_\odot-\Big(\frac{z}{z_0}\ddot{\vec r}^\perp_{\tt hs}+2\frac{v_{\tt sc}}{z_0}\dot{\vec r}^\perp_{\tt hs}\Big).
\label{eq:hs-perp-k-a}
\end{eqnarray}

\begin{figure}[t!]
\vskip 5pt
 \begin{center}
 \rotatebox{90}{\hskip 45pt  Acceleration, \,[$10^{-6}~{\rm m/s}^2$]}
\hskip -7pt
\begin{minipage}[b]{0.39\linewidth}
\rotatebox{0}{\hskip 120pt  $z_0=10~${\rm ly}}
\vskip -16pt
 \includegraphics[width=0.85\linewidth]{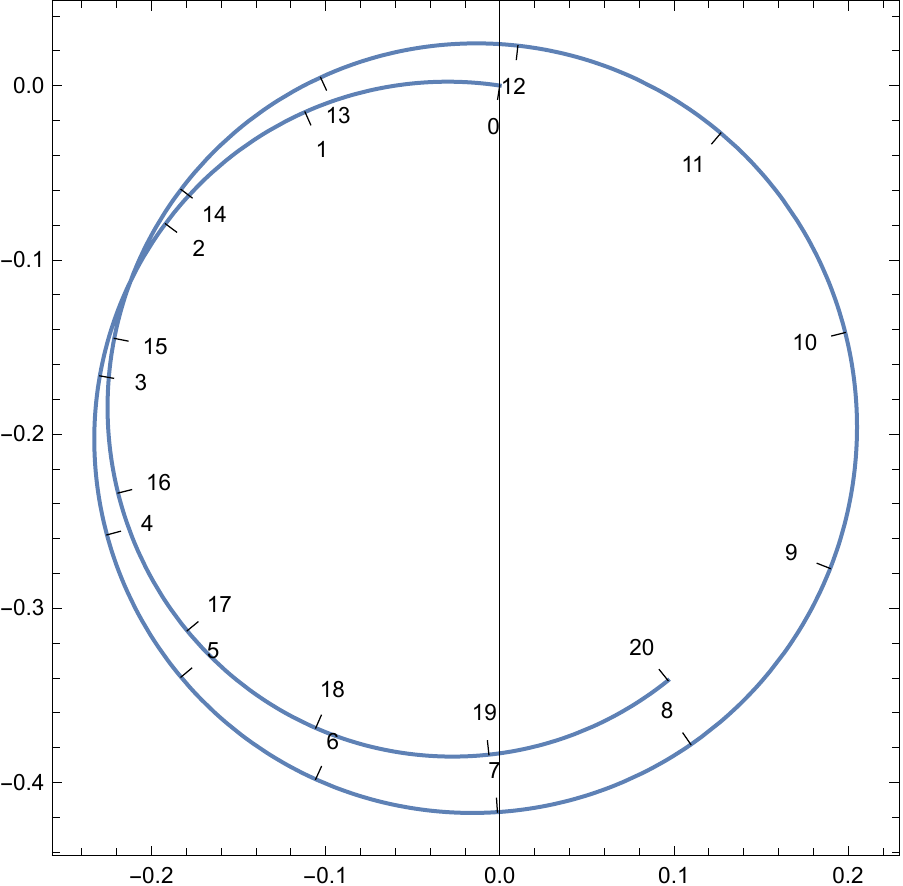}
 \rotatebox{0}{\hskip 22pt  Acceleration, \,[$10^{-6}~{\rm m/s}^2$]}
\end{minipage}
  \hskip 10pt
   \rotatebox{90}{\hskip 40pt  Acceleration, \,[$10^{-6}~{\rm m/s}^2$]}
\hskip -7pt
\begin{minipage}[b]{0.35\linewidth}
\rotatebox{0}{\hskip 100pt  $z_0=10~${\rm ly}}
\vskip -16pt
 \includegraphics[width=0.84\linewidth]{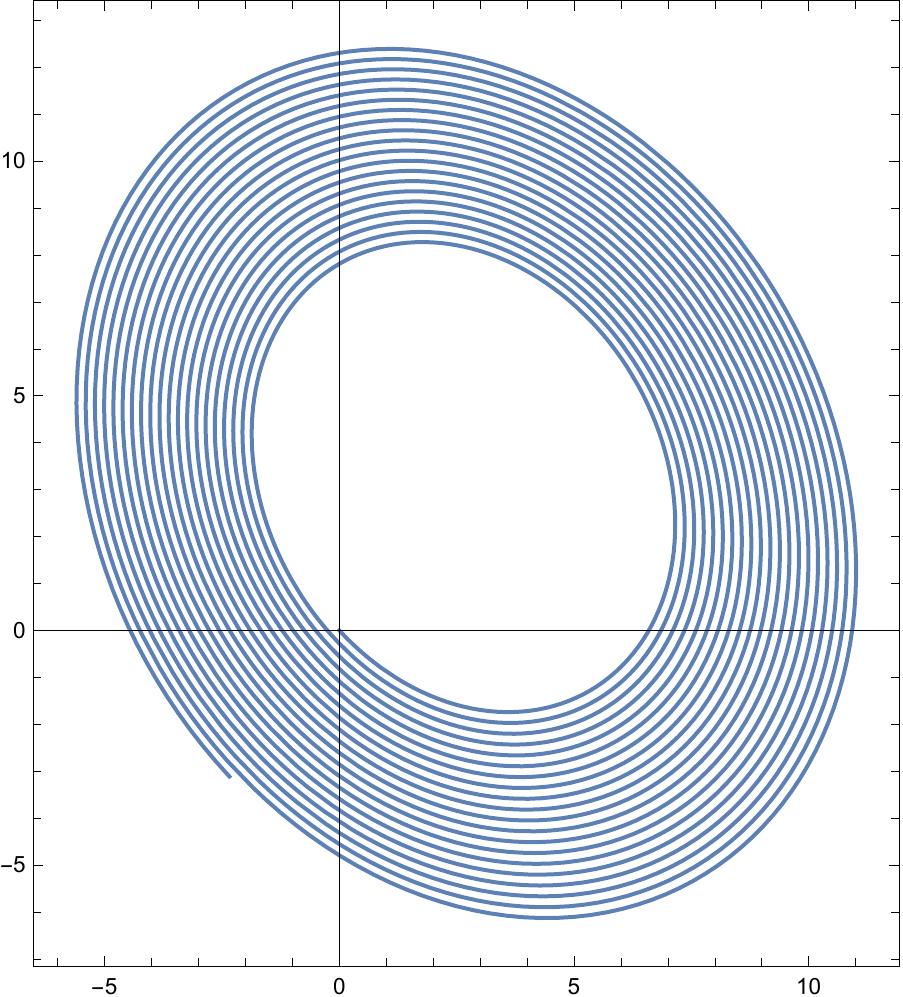}
 \rotatebox{0}{\hskip 20pt   Acceleration, \,[$10^{-6}~{\rm m/s}^2$]}
\end{minipage}
\vskip -5pt
\caption{\label{fig:acc-HS-Sun-10pc-ind} Acceleration of the primary optical axis for a target at $z_0=10$~ly and projected on the image plane moving at $v_{\tt sc}=25$~AU/yr. Left: Change in the acceleration (from its initial value at $z=550$~AU) due to reflex motion of the Sun. Ticks on the curves are in years. Right: change in the acceleration due the exoplanet, referred to its value at 550 AU.
}
\vskip 16pt
 \rotatebox{90}{\hskip 50pt  Acceleration, \,[$10^{-6}~{\rm m/s}^2$]}
\hskip -24pt
\begin{minipage}[b]{0.42\linewidth}
\rotatebox{0}{\hskip 100pt  $z_0=10~${\rm ly}}
\vskip -16pt
 \includegraphics[width=0.75\linewidth]{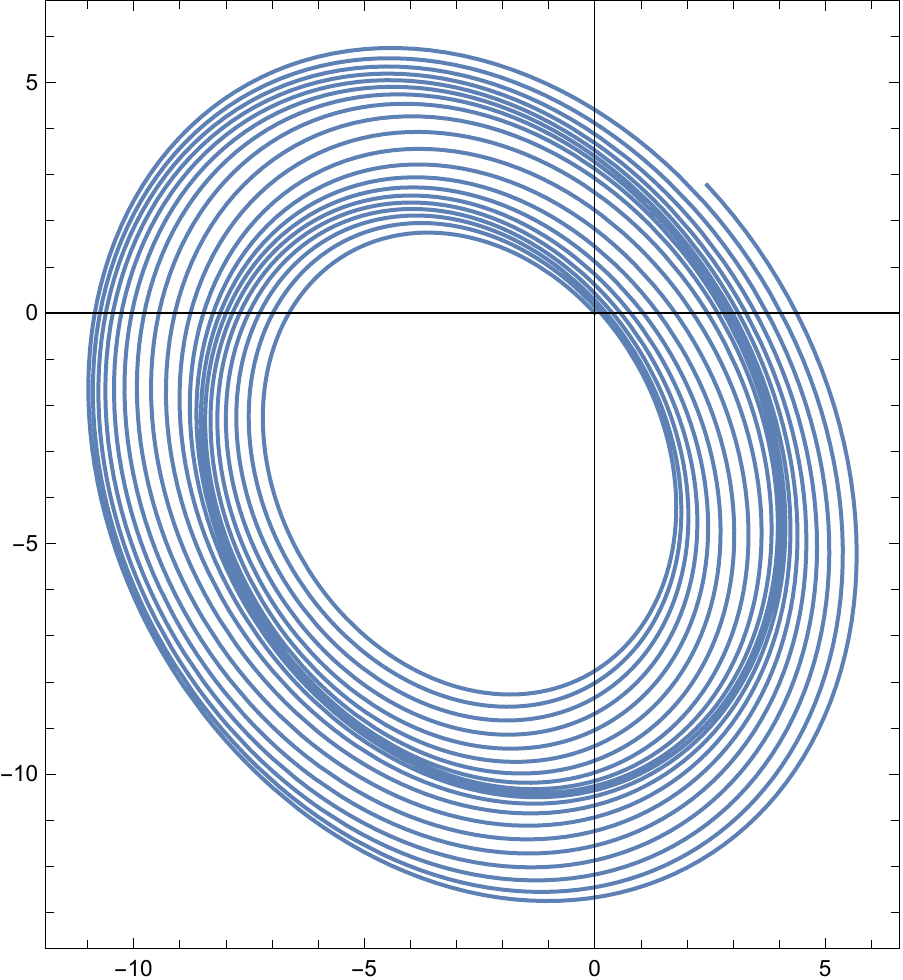}
 \rotatebox{0}{\hskip 20pt  Acceleration, \,[$10^{-6}~{\rm m/s}^2$]}
\end{minipage}
\vskip -5pt
\caption{\label{fig:acc-HS-Sun-10pc-comb} Combined change in the acceleration (referenced to its initial value) due to reflex motion of the exoplanet and the Sun, as modeled by (\ref{eq:pl-perp-k-a}) with individual components plotted in Fig.~\ref{fig:acc-HS-Sun-10pc-ind}.
}
\vskip 16pt
 \rotatebox{90}{\hskip 23pt  Acceleration, \,[$10^{-9}~{\rm m/s}^2$]}
\hskip -8pt
\begin{minipage}[b]{0.42\linewidth}
\rotatebox{0}{\hskip 140pt  $z_0=10~${\rm ly}}
\vskip -16pt
 \includegraphics[width=0.90\linewidth]{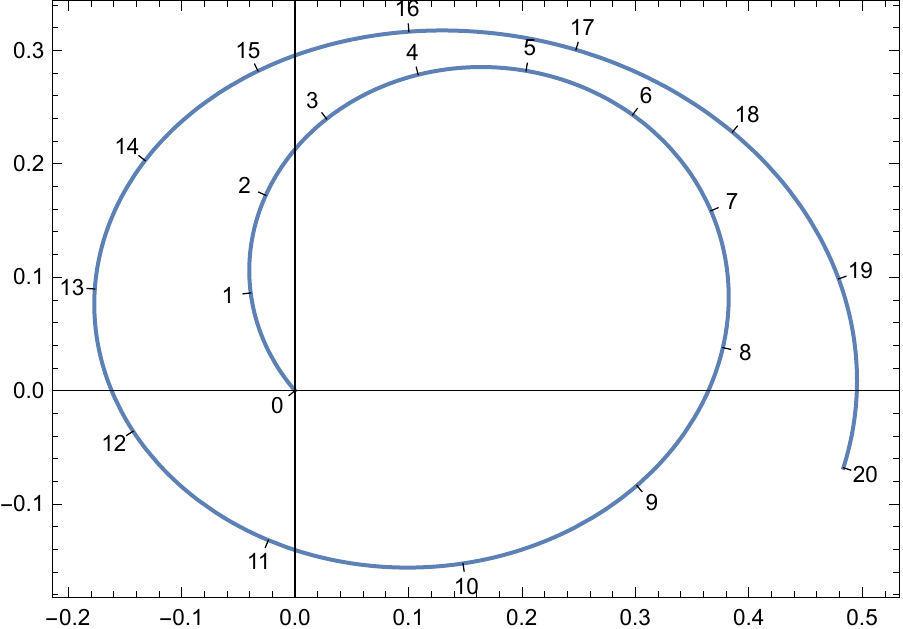}
 \rotatebox{0}{\hskip 23pt  Acceleration, \,[$10^{-9}~{\rm m/s}^2$]}
\end{minipage}
\vskip -5pt
\caption{\label{fig:acc-HS-10pc} Change in the acceleration (referenced to its initial value) due to reflex motion of the host star for a target at $z_0=10$~ly which is projected on a moving image plane with $v_{\tt sc}=25$ AU/yr, starting at $z=550$~AU. Ticks on the curves are in years.
}
 \end{center}
\end{figure}

To determine the acceleration vector due to the reflex motion of the Sun, $\ddot {\vec r}_\odot(t)$, we use the result (\ref{eq:pos-Xt-ddot}), that yields
{}
\begin{eqnarray}
\ddot {\vec r}_\odot(t) &=&
\sum_j \frac{m_j}{M_j} \frac{n_j^2a_j}{(1-e_j\cos E_j(t))^3}\bigg\{
  \left( \begin{aligned}
A_j& \\
B_j& \\
C_j&\\
  \end{aligned} \right)
\Big(\cos E_j(t) -e_j\Big)+
        \left( \begin{aligned}
F_j & \\
G_j & \\
H_j&\\
  \end{aligned} \right) \sqrt{1-e_j^2}\sin E_j(t)\bigg\}.~~~~~
  \label{eq:pos-mXt_sva}
\end{eqnarray}
Clearly, the expression for the host star, $\ddot {\vec r}_{\tt hs}(t) $, has the same form as (\ref{eq:pos-mXt_sva}), where one changes $\{A_j...H_j\} \rightarrow \{A^{\tt exo}_j...H^{\tt exo}_j\}$.  To derive this results, we assumed that the rate in the proper motion, $\dot\mu$, and any variability in the radial velocity, $\dot v_r$, are small for the duration of the science mission.

Similarly, using (\ref{eq:pos-Xt-ddot}), we derive the  expression for the acceleration vector of the exoplanet:
{}
\begin{eqnarray}
\ddot {\vec r}_{\tt p}(t) &=&   -
\frac{m_\odot}{m_\odot+m_{\tt E}}\frac{n^2_{\tt E}a_{\tt E}}{(1-e_{\tt E}\cos E_{\tt p}(t))^3}\bigg\{
  \left( \begin{aligned}
A^{\tt exo}_{\tt E}& \\
B^{\tt exo}_{\tt E}& \\
C^{\tt exo}_{\tt E}&\\
  \end{aligned} \right)
\Big(\cos E_{\tt E}(t) -e_{\tt E}\Big)+
        \left( \begin{aligned}
F^{\tt exo}_{\tt E} & \\
G^{\tt exo}_{\tt E} & \\
H^{\tt exo}_{\tt E}&\\
  \end{aligned} \right) \sqrt{1-e_{\tt E}^2}\sin E_{\tt E}(t)\bigg\}.~~~~~
  \label{eq:pos-vxaP}
\end{eqnarray}

We use the results (\ref{eq:pos-mXt_sva})--(\ref{eq:pos-vxaP}) to evaluate the acceleration that is experienced by the primary optical axis.

Fig.~\ref{fig:acc-HS-Sun-10pc-ind} shows the change in acceleration (from the initial value) due to reflex motion of the Sun (left), $\Delta \ddot {\vec r}^\perp_\odot$, and orbital motion of the exoplanet (right), $\Delta\ddot {\vec r}^\perp_{\tt p}$, that, from (\ref{eq:pl-perp-k-a}), have the form
 {}
\begin{eqnarray}
\Delta \ddot {\vec r}^\perp_\odot&=&\Big(1+\frac{z}{z_0}\Big)\ddot {\vec r}^\perp_\odot+ 2\frac{v_{\tt sc}}{z_0}\dot{\vec r}^\perp_\odot,
\qquad
\Delta\ddot {\vec r}^\perp_{\tt p}=-\Big(
\frac{z}{z_0}\ddot{\vec r}^\perp_{\tt p}+2
\frac{v_{\tt sc}}{z_0}\dot{\vec r}^\perp_{\tt p}\Big).
\label{eq:sun-pl-perp-k-ad}
\end{eqnarray}
Specifically, the left plot shows the induced acceleration of the optical axis (referenced to its initial value) due to the reflex motion of the Sun, $\Delta \ddot {\vec r}^\perp_\odot$. Ticks on the curve are in years, assuming that the image plane moves with velocity of $v_{\tt sc}=25$ AU/yr.
 The right plot presents the projected acceleration of the optical axis (referenced to its initial value) due to exo-Earth's orbital motion (also, at $z_0=10$~ly), $\Delta\ddot {\vec r}^\perp_{\tt p}$ and projected on the moving image plane. The inner ellipse corresponds to year 1 at $z=550$ AU, increasing outwards for 20 years. Again, the first terms in both expressions (\ref{eq:sun-pl-perp-k-v}) dominate, namely the $\ddot{\vec r}^\perp_\odot$ and $({z}/{z_0})\ddot{\vec r}^\perp_{\tt p}$, thus driving the mission requirements.

To appreciate the complexity of the relevant dynamics in the acceleration space, Fig.~\ref{fig:acc-HS-Sun-10pc-comb} shows combined change in the acceleration (referenced to its initial value) due to reflex motion of the exoplanet and the Sun, as modeled by (\ref{eq:pl-perp-k-a}) with individual components shown in Fig.~\ref{fig:acc-HS-Sun-10pc-ind}. Clearly, the dynamics of the exoplanet positioned at 10 ly dominates the solar reflex motion. For a more distant planet, the solar reflex motion will set a target-independent requirement for the acceleration sensitivity for the mission design.

Finally, Fig.~\ref{fig:acc-HS-10pc} shows the acceleration of the primary optical axes induced by the reflex motion of the host star, $\Delta \ddot {\vec r}^\perp_{\tt hs}$, that from (\ref{eq:hs-perp-k-a}) has the same form as $\Delta\ddot {\vec r}^\perp_{\tt p}$ in (\ref{eq:sun-pl-perp-k-ad}) where ${\vec r}^\perp_{\tt p}$ is replaced with ${\vec r}^\perp_{\tt hs}$. Note that, similar to the case of host star's  velocity, this acceleration is a factor of $\sim10^3$ smaller than that of the Sun's and even smaller than that of the exoplanet and, thus, is not a significant requirement driver.

We note that because of the telescope's egress from the solar system, the acceleration that is needed to follow the image increases. Assuming the maximum acceleration of the primary optical axis to be similar to that of the projected velocity of the exoplanet, $a_0=6.1\times 10^{-6}~{\rm m/s}^2$, we estimate the needed $\Delta v$ for $\Delta t=20$~years of operations as
$\Delta v=a_0 \Delta t \sim 3.78$~km/s, which is the value achievable with modern electric micropropulsion systems.

\section{Toward imaging with the SGL}
\label{sec:RF-and-imaging}

\subsection{Prior knowledge about the exoplanet}
\label{sec:a-priory}

With the estimates of the dominant effects in the motion of the primary optical axes on the image plane, we now need to put all this information in the context of a realistic mission. Given the current state of the propulsion techniques, an SGL telescope is thought to be  a single-planetary-system instrument, able to explore all planets and satellites in that system. In this case, the target must be well-justified. Evaluating what we may already know about the exoplanet (e.g., rotation period, prevalence of clouds) will be important for establishing the mission requirements, optimizing the reconstruction of a spatially resolved image, and motivating precursor projects.

The occurrence rate of Earth-sized terrestrial planets in the habitable zones (HZs) of Sun-like (FGK) stars remains a much-debated quantity.  Only a handful of such planets have been discovered.  Current estimates range from 2\%  to 22\%.  The SIMBAD Astronomical Database\footnote{For details on the SIMBAD Astronomical Database, see {\tt http://simbad.u-strasbg.fr/simbad/.}} lists 8589 F stars, 5309 G stars, and 1688 K stars within $\sim$ 100 ly.  Taking even the lowest estimates, we can expect to detect at least one terrestrial planet in the HZ of a star within 100 ly in the near future.  Once such a planet is discovered significant observational resources will be devoted to characterizing it.

Most likely, we will want to image Earth 2.0, around a G-type star (some of these stars are listed in Table~\ref{tb:pdreceive2}), which is not transiting.  A imaging mission to the focal region of the SGL could follow a ``big terrestrial planet finder'' that observes an exo-Earth around a G star and measures its spectra. We should be very confident that the selected target is habitable or even inhabited. A telescope at the SGL would be the next major step, possibly the biggest step in the 21st century for exoplanet exploration.

If the planetary atmosphere contains oxygen and, possibly, signs of life, the next step would be to launch the SGL focal mission to image this planet at high resolution. The planet's orbit would have to be measured in 3-D, using either astrometry and/or RV measurements combined with direct imaging. If we are lucky, it will be inclined so that it transits, providing a radius. These measurements would allow us to obtain the information and point the telescope.

Once we know of a terrestrial planet in a HZ so close to our own, we posit that significant resources will be devoted to characterizing the planet and its system using the above techniques. The knowledge we gain from this will include:
\begin{inparaenum}[i)]
\item orbital ephemeris, to at least milliarcsecond accuracy and precision;
\item detailed knowledge of the atmosphere, including temperature, structure, chemical composition, and albedo, all inferred from non-spatially-resolved spectroscopy;
\item estimates of rotation rate, gained from temporal monitoring of the spectroscopy; and
\item some understanding of cloud and surface properties from Doppler imaging.
\end{inparaenum}

A mission to the SGL focal region would begin after the discovery of an exo-Earth, and there would be $\sim22$ years of ``cruise'' before the telescope would reach 547 AU, moving with $\sim25$~AU/yr. During those 20+ years, the parent star's location would have been observed with 1~$\mu$as precision at least 100 times, so that its position would be known at the $0.1~\mu$as level. The parent star's position would be known to $\sim45$~km at 30 pc. The orbit period of the planet would be known to $<1\%$ meaning that the semi-major axis is known to $\sim0.7\%$ or $\sim $1 million km.  If the planet is in a face-on orbit, we will know the radial distance to $\sim1$ million km, but the error bar in the tangential direction will be a factor of $\sim 6$ larger. The diameter of the Earth-like planet is $\sim13,000$ km, so that the area on the sky we must search is an ($80 \times 500$) grid. Once the SGL focal mission's telescope detects the planet, it would scan a much smaller area to define the ``edges'' of the planet. Astrometric observations of the star when the planet was discovered would have measured its mass, that plus its size would give us the density of the planet.

\subsection{Local image-centric reference frames}
\label{sec:loc-RefFrames}

The knowledge of the dynamics of the primary optical axis of the host star in the focal region of the SGL may be used to establish the host star image-centric local reference frame, as discussed in Sec.~\ref{sec:definition}. Such an image-centric coordinate reference frame may be realized using image sensor data from the imaging telescope. This is enabled by the fact the self-luminous sources result in very bright Einstein rings formed around the Sun.

In Refs.~\cite{Turyshev-Toth:2019-blur,Turyshev-Toth:2019-image,Turyshev-Toth:2019-extend}  we developed a comprehensive wave-optical approach to describe gravitational lensing phenomena (summarized in Sec.~\ref{sec:aintro}). For a self-luminous resolved source with luminosity of $L_{\tt s}$, the power density, $B_{\tt lum}$, received on the image plane  in the strong interference region of the SGL is given as
{}
\begin{eqnarray}
B_{\tt SGL}= \frac{ L_{\tt s}}{4\pi R_{\tt s}^2} \frac{2R_{\tt s}}{z_0}\sqrt{\frac{2r_g}{ z}}
~~~{\rm W}/{{\rm m}^2},
  \label{eq:Pexo-pdsl}
\end{eqnarray}
where $R_{\tt s}$ its radius, and $z_0$ its distance from the Sun. The same quantity for a direct observation without SGL is
{}
\begin{eqnarray}
B_{\tt 0}= \frac{L_{\tt s}}{4\pi R_{\tt s}^2}  \Big(\frac{R_{\tt s}}{z_0}\Big)^2~~~{\rm W}/{{\rm m}^2}.
  \label{eq:Pexo-pdsl0}
\end{eqnarray}

Using the method presented in \cite{Turyshev-Toth:2020-extend}, one may estimate the photon power density that may be received from a host stars on the image plane at the strong interference region of the SGL. Table~\ref{tb:pdreceive2} provides the estimates for some of the ``solar analogs'' in our stelar neighborhood as they would be observed at the heliocentric distance of 650 AU.  Also, according to (\ref{eq:S_z*6z-pos}), we have $r_{\tt s}=(z/z_0) R_{\tt s} $ being the size of the image.  Note that the amplified light of the host star will be very bright, overshigning that of the solar corona. Such observations will still require a coronagraph, but not to the same level of rejection. These estimates suggest that we may use this light from the host star to study the exoplanet before it will have the SNR large enough to be observed on the background of the solar corona. We note that the power density in the image plane for these solar analogs is very similar, $B_{\tt SGL}={\cal O}(10^{-6})$~W/m$^2$. As shown in (\ref{eq:Pexo-pdsl}), it depends only linearly on the inverse of the heliocentric distance of the source, $z_0$.

\begin{table*}[t!]
\vskip-15pt
\caption{Power density from some of the nearest ``solar analogs'', received in the image plane at the strong interference region of the SGL at the heliocentric distance of 650 AU. Data from \protect\url{https://en.wikipedia.org/wiki/Solar_analog}
\label{tb:pdreceive2}}
\begin{tabular}{|l|c|c|c|c|c|c|}\hline
Source  &$L_{\tt s}$, in $L_\odot$ & $z_0$, ly &  $R_{\tt s}$, in $R_\odot$ &  $B_0$, ${\rm W/m}^2$ & $r_{\tt s}$, km& $B_{\tt SGL}$, ${\rm W/m}^2$  \\\hline\hline
Sun & 1\phantom{.0000} & $-$  & 1\phantom{.0000} & $3.28 \times 10^{-3\phantom{0}}$ & $-$ & $-$\\
Solar corona & $5\times 10^{-5}$ & $-$  & $-$ & $1.07 \times 10^{-9\phantom{0}}$ & $-$ & $-$\\
\hline\hline
Sun-like star at 10 ly& 1\phantom{.0000} & 10.0  & 1\phantom{.0000} & $3.40 \times 10^{-9\phantom{0}}$ & 715 & $7.21 \times 10^{-6}$\\
Sun-like star at 100 ly& 1\phantom{.0000} & 100.0\phantom{0}  & 1\phantom{.0000} & $3.40 \times 10^{-11}$& 71.5  & $7.21 \times 10^{-7}$\\
\hline\hline
Sigma Draconis         & 0.41\phantom{00} & 18.8 & 0.776\phantom{0} & $3.95\times 10^{-10}$ &	295 & $2.03\times 10^{-6}$\\
Beta Canum Venaticorum & 1.151\phantom{0} & 27.4 & 1.123\phantom{0} & $5.22\times 10^{-10}$ &	293 & $2.70\times 10^{-6}$\\
61 Virginis            & 0.8222           & 27.8 & 0.9867           & $3.62\times 10^{-10}$ &	254 & $2.16\times 10^{-6}$\\
Zeta Tucanae           & 1.26\phantom{00} &	28.0 & 1.08\phantom{00} & $5.47\times 10^{-10}$ &	276 & $3.00\times 10^{-6}$\\
Beta Comae Berenices   & 1.357\phantom{0} & 29.8 & 1.106\phantom{0} & $5.20\times 10^{-10}$ &	266 & $2.97\times 10^{-6}$\\
61 Ursae Majoris       & 0.609\phantom{0} & 31.1 & 0.86\phantom{00} & $2.14\times 10^{-10}$ &	198 & $1.64\times 10^{-6}$\\
HR 511                 & 0.516\phantom{0} & 32.8 & 0.819\phantom{0} & $1.63\times 10^{-10}$ &	179 & $1.38\times 10^{-6}$\\
Alpha Mensae           & 0.832\phantom{0} & 33.1 & 0.99\phantom{00} & $2.58\times 10^{-10}$ &	214 & $1.83\times 10^{-6}$\\
HD 69830               & 0.622\phantom{0} & 40.6 & 0.905\phantom{0} & $1.28\times 10^{-10}$ &	160 & $1.22\times 10^{-6}$\\
HD 10307               & 1.44\phantom{00} & 41.2 & 1.14\phantom{00} & $2.89\times 10^{-10}$ &	198 & $2.21\times 10^{-6}$\\
HD 147513              & 0.98\phantom{00} & 42.0 & 1.0\phantom{000} & $1.89\times 10^{-10}$ &	170 & $1.68\times 10^{-6}$\\
58 Eridani             & 0.992\phantom{0} & 43.3 & 0.96\phantom{00} & $1.80\times 10^{-10}$ &	159 & $1.72\times 10^{-6}$\\
47 Ursae Majoris       & 1.48\phantom{00} & 45.9 & 1.172\phantom{0} & $2.39\times 10^{-10}$ &	183 & $1.98\times 10^{-6}$\\
Psi Serpentis          & 0.98\phantom{00} & 47.8 & 1.0\phantom{000} & $1.46\times 10^{-10}$ &	150 & $1.48\times 10^{-6}$\\
20 Leonis Minoris      & 1.378\phantom{0} & 49.1 & 1.247\phantom{0} & $1.95\times 10^{-10}$ &	182 & $1.62\times 10^{-6}$\\
Nu Phoenicis           & 2.0\phantom{000} & 49.3 & 1.26\phantom{00} & $2.80\times 10^{-10}$ &	183 & $2.32\times 10^{-6}$\\
51 Pegasi              & 1.36\phantom{00} & 50.9 & 1.237\phantom{0} & $1.79\times 10^{-10}$ &	174 & $1.56\times 10^{-6}$\ \\\hline\hline
\end{tabular}
\end{table*}

\begin{figure}[h!]
\begin{tabular}{ccccc}
$\rho=5r_{\tt s}$&$\rho=2r_{\tt s}$&$\rho=1.25r_{\tt s}$&$\rho=r_{\tt s}$&$\rho=0$\\
~&~&~&~&~\\
\includegraphics{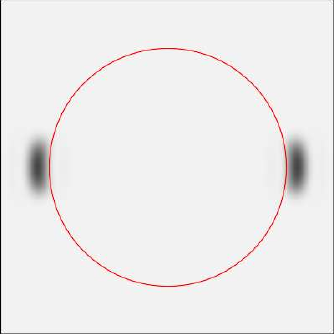}&\includegraphics{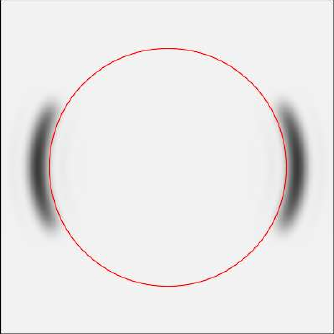}&\includegraphics{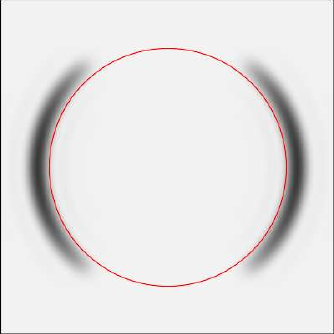}&\includegraphics{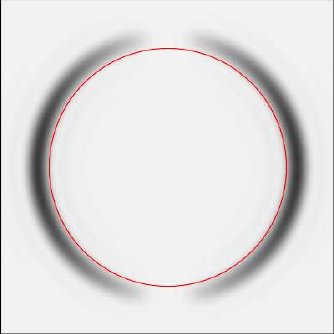}&\includegraphics{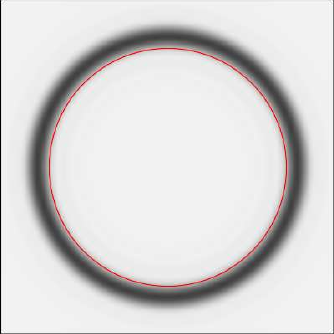}
\end{tabular}
\caption{\label{fig:SGL-im-sensor} Evolution of the Einstein ring formed around the image of the Sun (represented by a red circle) on the image sensor in the focal plane of the imaging telescope. (The behavior of the ring is given as a function of the distance from the optical axis, $\rho$. Shown for a Sun-like star at 10~ly and the telescope at 650~AU, thus the size of the host star image is $r_{\tt s}\sim 715$~km.)
The orientation of the primary and secondary images of the star and their evolution into an Einstein-ring, as seen by the telescope approaching the host star optical axis, indicate the direction towards the center of the star's projected image in the image plane, while the brightness and angular extent of the arcs seen by the telescope indicate distance. Additional spacecraft can help to greatly improve navigational accuracy by triangulating.
}
\end{figure}

As seen in Table~\ref{tb:pdreceive2}, the brightness of the signals received from the stars in our neighborhood is much higher that that of the solar corona. The SGL itself therefore provides us with a means to find the position of the host star image. Evolving morphology of the Einstein ring corresponding to a host star signal received by the image sensor in the focal plane of an imaging telescope may be used to our advantage.  As discussed in \cite{Turyshev-Toth:2019-image,Turyshev-Toth:2020-extend}, a multipixel image sensor provides a very sensitive means to determine position. For any extended source, this can be done by monitoring the formation of the Einstein ring on the image sensor as a function of the distance from the optical axis, $\rho$. As shown in Fig.~\ref{fig:SGL-im-sensor}, this process includes several critical phases that begin when the telescope is far away from the primary optical and is outside the physical extent of the image. At this point, only one bright spot corresponding to the major image is present on the sensor. As the telescope moves closer to the optical axis, the minor image begin to emerge from behind the Sun thus resulting in two spots of increasing brightness that correspond to the major and minor images. A line drawn from the major to minor image will point toward the optical axis. The relative brightness and morphology of the two images may be used to determine the distance to the primary optical axis. Positions of the two spots along the yet-to-be formed Einstein ring, provide angular information for the primary optical axis in the telescope-centric coordinate reference frame.

As the telescope moves closer to the optical axis, the spots evolve into two arclets, then into larger arcs, which eventually merge to form a complete Einstein ring. At this point the telescope enters the physical extent of the image. This motion culminates in the fully formed Einstein ring appearing on the sensor at its maximum brightness. This happens when the telescope is positioned exactly at the center of the projected image of the star, i.e., on the primary optical axis. The evolving range data (i.e., relative brightness) and associated angular information allow for a precise determination of the telescope's position with respect to the primary optical axis in real time. Monitoring this well-understood process forms a method of guidance that may be used to establish a local reference frame in the focal region of the SGL.

Following this process, the exoplanet image located $\sim10^4$~km from the host star image in the image plane. The approximate direction to the exoplanet is known, which aids in the search. Approaching the exoplanet image is a repeat of the host star approach but on a smaller and fainter scale, on the host star light background. (Quantifying relative brightness is a work-in-progress. Results, when available will be presented elsewhere.) The next step is to move the spacecraft with the exoplanet image size in a meter-scale precision.

Precise monitoring of the amplified light from the host star may be used to find the exoplanet as this light is so bright that even minute changes in the host star position will be detectable.  So, finding a much dimmer Earth-like exoplanet will not be a significant challenge. Depending on the brightness of the exoplanetary signal, the size of the imaging telescope and the performance of the coronagraph may be used to establish the exoplanet image-centric local reference frame (as discussed in Sec.~\ref{sec:definition}) following the same process  as for the host star (i.e., when we rely on the readings from the image sensor on the focal plane of the imaging telescope.) Alternatively, we may establish and maintain a tight formation of imaging telescopes, while precisely monitoring their relative positions. This relative position information may be used as a grid with the center that may be treated as a solve-for parameter during the image deconvolution process  \cite{Turyshev-Toth:2020-extend,Toth-Turyshev:2020}.

Once the exoplanet image is located, two or more telescopes could establish its precise boundary and track its motion. These telescopes may simultaneously participate in the observational campaign and participate in establishing and maintaining a local reference frame. The observing telescope(s) may move from pixel to pixel on a meter-scale grid with respect to this frame, following a pattern that minimizes velocity changes while maximizing science return.

After the planet is located, the observational mission begins. During the mission, the observing telescope must track the complex motion of the planet's projected image in the SGL image plane. Without acceleration, the observing telescope would follow an inertial trajectory egressing the solar system, a high-speed hyperbolic escape trajectory in the SSB reference frame. The exoplanetary image, meanwhile, undergoes complex motion in the image plane as a result of all the effects that we have accounted for in this paper, including solar reflex motion and the exoplanet's own orbital motion, as well as other smaller effects. Together, these determine requirements on the SGL mission's propulsion capability, which is needed in order to ensure that the observing telescope can maintain its position within the projected image of the exoplanet that it samples.

The ultimate imaging would have to be done by positioning the telescope within the image at meter-class relative precision. This precision is especially important in velocity space. A displacement in the image plane simply results in a shifted image of the exoplanet; so long as this shift is modest, the exoplanet remains within the image area. Uncontrolled velocity differences, however, will result in the smearing of image pixels, degrading the quality of observation and ultimately rendering image reconstruction impossible.

The process outlined have allows for improved navigational precision for the ultimate exoplanet imaging mission to the SGL focal region that may be summarized as follows:  i) Coarse-grade navigation done with $\sim100,000$~km precision needed to find the region where host star light amplification is observable; ii) Medium-grade navigation done with $\sim100$~km precision needed to find the host star image and navigate toward the exoplanet image; iii) Fine-grade navigation done with $\sim1$~km precision needed to find the exoplanet image and establish local reference frame; and iv) Conduct imaging observations with $\sim1$~m precision that is needed to establish relative position within the exoplanetary local reference frame for pixel-by-pixel data collection.

\section{Discussion and Conclusions}
\label{sec:disc}

In previous studies of the SGL, we considered a static configuration: A motionless exoplanet's image, projected onto an image plane by the motionless Sun, to be sampled by a roving telescope that measures the intensity of light from the exoplanet while traversing a kilometer-scale projection in the image plane.

Reality is more complicated. Even if considering only first order effects (i.e., omitting planetary interactions and the resulting perturbations), we need to account for a complex set of motions. First, the lens itself, our own Sun, ``wobbles'', undergoing complex motion as a result of the gravitational pull of solar system planets, most notably Jupiter. The target exoplanet follows its own elliptical orbit around its host star. The host star itself wobbles as a result of the gravitational pull of the planets in that system. Lastly, the exoplanetary system may have proper motion relative to our own solar system, as the two systems follow different galactocentric orbits and do not remain at rest relative to each other.

Consequently, the image of the target exoplanet, as projected into a hypothetical image plane in the SGL focal region, undergoes similarly complex noninertial motion. A telescope, placed in the image plane and used for a sustained imaging campaign to obtain a high resolution, multispectral image of the target exoplanet, must be navigated precisely within the image plane. Its position and velocity relative to the projected exoplanetary image must be known very accurately at all times, and controlled accurately as well.

How accurately? A projected exoplanetary image for prospective imaging targets may be anywhere between less than a kilometer to several kilometers in size, depending on the size of the exoplanet, its distance from the Sun, and the distance of the imaging telescope from the Sun. Dividing this image into several thousand (up to a million) image pixels implies a pixel size measured in meters. Incorrect or insufficient knowledge of the position of the imaging telescope can smear its observations in unpredictable ways as the telescope inadvertently traverses across image pixels, introducing uncontrolled noise that would significantly impact or undermine image reconstruction.

Fortunately, it is not necessary to know the exact position of the imaging instrument with meter-scale precision in the SSB reference frame. It is sufficient to know its position relative to the projected exoplanet image. The location of this image can be well established using {\em in situ} measurements by the observing telescopes in combination with an accurate model of all motions. This approach allows us to introduce a local reference frame, which can then form the basis for image reconstruction.

Accordingly, to facilitate imaging with the SGL, we introduced new references frames in the form of image-centric coordinates that rely on the source's light amplified by the SGL and projected into an image plane in the SGL focal region. These frames is mission-enabling, allowing for high-precision imaging operations within the SGL focal region.

Before we can study the steps required to reliably establish such a local reference frame, however, we need to understand the full range of motions that the projected image exhibits. This knowledge is required in order to evaluate the very feasibility of placing a telescope in the SGL focal region for exoplanet imaging, and will help with establishing a baseline for actual mission requirements, in terms of acceleration and acceleration accuracy.

For the fictitious target system that we investigated (see  Table~\ref{tab:ssplanets-exo}), the characteristic acceleration of the projected exoplanet image is $\sim 6~\mu$m/s$^2$. An actual imaging mission may rely on continuous acceleration of this magnitude, or it may utilize a different strategy (e.g., inertial motion with intermittent trajectory changes to maximize positioning accuracy, optimize the use of consumables and the propulsion system, perhaps even limit the release of contaminating exhaust into the immediate vicinity of the observing telescope). Such technical details are yet to be determined and are beyond the scope of our present analysis.

Our results also help us understand the cumulative velocity changes during an imaging mission. This is critical information for mission design, as it directly translates into propulsion system requirements and consumables (fuel) over a multiyear mission. For the fictitious target system that was used in our analysis, even an extended science mission of 20 years in duration will require a $\Delta v$ capability of less than 4~km/s. The actual value, of course, depends on the specific target, as well as the navigational strategy that is used in the image plane to optimize the science mission, which is yet to be determined. Once a specific target is identified, the corresponding velocity changes required to track its image will be a key determining factor for the maximum duration of an imaging mission and the resulting achievable science objectives.

It was with these considerations in mind that we evaluated the changes in position, the velocity, and the acceleration of the projected image of a realistic, albeit hypothetical, exoplanet in the image plane. These results can be used directly in order to establish an optimal design for an observational campaign, which may use one or several instruments and continuous or intermittent acceleration, to achieve the best possible result for robust image recovery.

Our conclusion is that exoplanet imaging via the SGL requires several key technologies that are challenging, such as the determination of an exoplanet orbit to a meter class relative precision and the motion and stabilization of the spacecraft over millions of pointings with limited power. However, these challenges, though substantial, are not insurmountable. We are addressing these issues systematically as part of our ongoing effort to conceptualize a technologically feasible, scientifically valuable SGL mission. Results, when available will be published elsewhere.

\begin{acknowledgments}
We thank Thomas Heinsheimer, Henry Helvajian, Darren Garber, Artur Davoyan, John McVey, and Louis Friedman for their comments, suggestions, and constructive criticisms on this manuscript. This work in part was performed at the Jet Propulsion Laboratory, California Institute of Technology, under a contract with the National Aeronautics and Space Administration.
VTT acknowledges the generous support of Plamen Vasilev and other Patreon patrons.

\end{acknowledgments}


\appendix

\section{Position, velocity and acceleration for a given set of orbital elements}
\label{sec:Kepler}

\subsection{The Keplerian problem}

We present the position vector of the orbit as
{}
\begin{eqnarray}
\vec r&=& r\vec n_r,
\label{eq:pos-vec-r}\\
\dot {\vec r}&=& \dot r\vec n_r +r\dot \theta \vec n_\theta,
\label{eq:pos-vec-rdot}\\
\ddot {\vec r}&=& (\ddot r-r\dot \theta^2)\vec n_r +\Big[\frac{1}{r}\frac{d}{dt}(r^2\dot \theta)\Big] \vec n_\theta.
\label{eq:pos-vec-rddot}
\end{eqnarray}

In the case of a binary Kepler problem, the angular momentum,
{}
\begin{eqnarray}
\vec h&=& r^2\dot \theta \vec n_z,
\label{eq:pos-vec-h}
\end{eqnarray}
is conserved, namely $h=r^2\dot \theta={\rm const}$. Another equation to describe the binary motion is
{}
\begin{eqnarray}
\ddot r-r\dot \theta&=& -\frac{G(m_1+m_2)}{r^2}.
\label{eq:pos-vec-rr}
\end{eqnarray}
Using these equations, we determine
{}
\begin{eqnarray}
r&=& \frac{p}{1+e\cos(\theta-\varpi)}, \qquad {\rm where}\qquad p=\frac{h^2}{G(m_1+m_2)}=a(1-e^2).
\label{eq:pos-vec-rpg}
\end{eqnarray}
So, as a result, we have:
{}
\begin{eqnarray}
r&=& \frac{a(1-e^2)}{1+e\cos(\theta-\varpi)} \qquad {\rm and,~defining,}\qquad f=\theta-\varpi, \qquad {\rm we~ obtain}
\label{eq:pos-vec-rffp1}
\end{eqnarray}
{}
\begin{eqnarray}
r&=& \frac{a(1-e^2)}{1+e\cos f}.
\label{eq:pos-vec-rffp}
\end{eqnarray}

We also need the quantities $\dot r$ and $r\dot f$ are the time rate of change of separation and angular distance from the focus of the ellipse. These quantities are given as
{}
\begin{eqnarray}
\dot r&=&\frac{na}{\sqrt{1-e^2}}e\sin f,
\label{eq:pos-r-dot}
\qquad
r\dot f=\frac{na}{\sqrt{1-e^2}}\Big(1+e\cos f\Big),
\label{eq:pos-rf-dot}
\end{eqnarray}
where $n=2\pi/T$ is the mean motion.

We note that only $r$, $\dot r$ and $f$ vary with time. The time-dependence is determined from Kepler's equation
{}
\begin{eqnarray}
\dot r&=& \frac{na}{r}\sqrt{a^2e^2-(r-a)^2},
\label{eq:pos-vec-K}
\end{eqnarray}
which is solved by introducing a new variable, $E$, the eccentric anomaly, by means of the substitution:
{}
\begin{eqnarray}
r&=& a(1-e\cos E).
\label{eq:pos-vec-E}
\end{eqnarray}
The differential equation (\ref{eq:pos-vec-K}) transforms to
{}
\begin{eqnarray}
\dot E&=& \frac{n}{1-e\cos E}.
\label{eq:pos-vec-E-dor}
\end{eqnarray}
The solution can be written as
{}
\begin{eqnarray}
n(t-t_0)&=& E-e\sin E,
\label{eq:pos-M0=}
\end{eqnarray}
where we have taken $t_0$ to be the constant of integration and used the boundary condition $E=0$ when $t=t_0$. At this point we can define a new quantity, $M$, the mean anomaly, such that
{}
\begin{eqnarray}
M&=& n\big(t-t_0\big) \qquad {\rm and} \qquad
M=E-e\sin E.
\label{eq:pos-M1}
\end{eqnarray}

This is a transcendental equation which must be solved numerically. Also, the orbital period, $T$, is derived from Kepler's 3rd law as
{}
\begin{eqnarray}
T^2=\frac{4\pi^2 a^3}{G(m_1+m_2)}.
\label{eq:pos-f}
\end{eqnarray}

\subsection{Modeling orbital positions}

Using (\ref{eq:pos-vec-rffp}), we introduce Cartesian components of the vector from (\ref{eq:pos-vec-r}) as
{}
\begin{eqnarray}
x&=& r\cos f \qquad {\rm and}\qquad y=r\sin f.
\label{eq:pos-vec-df}
\end{eqnarray}

Introducing Cartesian unit vectors along $\vec i=\hat{\vec x}$, $\vec j=\hat{\vec y}$ and $\vec k=\hat{\vec z}$,   we present (\ref{eq:pos-vec-r})  as
{}
\begin{eqnarray}
\vec r&=& x\vec i+y\vec j +0 \vec k.
\label{eq:pos-vec-r2}
\end{eqnarray}

At this point, we project position vector (\ref{eq:pos-vec-r2}) onto a plane of the sky by rotating it with respect to the angles $\Omega$ and $\omega$. As a result we derive the following expressions:
{}
\begin{eqnarray}
    \left( \begin{aligned}
x & \\
y & \\
z &\\
  \end{aligned} \right)
 &=&r
    \left( \begin{aligned}
\cos\Omega\cos(\omega+f)-\sin\Omega\sin(\omega+f)\cos i & \\
\sin\Omega\cos(\omega+f)+\cos\Omega\sin(\omega+f)\cos i &\\
\sin(\omega+f)\sin i &\\
  \end{aligned} \right).~~~~~
  \label{eq:pos-x}
\end{eqnarray}

In the expressions above, we use the coordinate system such that positive directions of the coordinate axis in the plane, $+x$ and $+y$ correspond to the observed  $+\Delta {\rm Dec}~(\Delta\delta)$ and  $+\Delta {\rm RA}~(\Delta\delta)$ respectively between the orbiting body and central object. In this coordinate system, the positive direction $+z$ is defined toward the observer, contrary to the more commonly used convention aligned with the radial velocity.

It is convenient to separate in (\ref{eq:pos-x}) the time-dependent terms and those that are time-independent. This may be done by using normalized Thiele-Innes constant elements (making them, essentially, to be the direction cosines):
{}
\begin{eqnarray}
A&=& \cos\Omega\cos\omega-\sin\Omega\sin\omega\cos i,
\label{eq:pos-A}\\
B&=& \sin\Omega\cos\omega+\cos\Omega\sin\omega\cos i,
\label{eq:pos-B}\\
F&=&-\cos\Omega\sin\omega-\sin\Omega\cos\omega\cos i,
\label{eq:pos-F}\\
G&=&-\sin\Omega\sin\omega+\cos\Omega\cos\omega\cos i,
\label{eq:pos-G}\\
C&=& \sin\omega\sin i,\\
\label{eq:pos-C}
H&=& \cos\omega\sin i.
\label{eq:pos-H}
\end{eqnarray}

We also define the quantities called the elliptical rectangular coordinates, {}
\begin{eqnarray}
X&=&\cos E-e,
\label{eq:pos-X}\\
Y&=& \sqrt{1-e^2}\sin E,
\label{eq:pos-Y}
\end{eqnarray}
so that $r^2=a^2(1-e\cos E)^2=a^2(X^2+Y^2),$ and $r\cos f =a X$, $r\sin f=a Y$.

Measured separations and position angels $(\rho,\phi)$ at the time $t$ are related to the projected quantities $(x,y,z)$ by simple relationships $\Delta \delta=x=\rho\cos\phi$ and  $\Delta \alpha\cos \delta=y=\rho\sin\phi$. Using the expressions  (\ref{eq:pos-A})--(\ref{eq:pos-H}), we may present (\ref{eq:pos-x}) as below:
{}
\begin{eqnarray}
   \left( \begin{aligned}
 x & \\
 y & \\
 z &
  \end{aligned} \right)
 &=&
  a\bigg\{
    \left( \begin{aligned}
A & \\
B & \\
C & \\
  \end{aligned} \right)  X(t)+    \left( \begin{aligned}
F & \\
G&  \\
H & \\
  \end{aligned} \right)  Y(t)\bigg\},~~~~~
  \label{eq:pos-Xt}
\end{eqnarray}
where the time-dependence is  only in $E(t)$ whose dependence determined from Kepler's equations  (\ref{eq:pos-M1}).

We also recognize that the instantaneous position on the sky of a star at time $t$ is determined by a fixed displacement $(\alpha_0,\delta_0)$ at reference time $t_0$, the proper motion vector $(\mu_\alpha,\mu_\delta)$ (where the proper motion component in the right ascension includes the $\cos\delta$ factor), and the orbital elements. In Cartesian coordinates laying on a plane tangent to the sky and directed along right ascension and declination, respectively, the position can be written as
{}
\begin{eqnarray}
    \left( \begin{aligned}
\Delta x & \\
\Delta y &
  \end{aligned} \right) =    \left( \begin{aligned}
z_0\Delta \delta(t) & \\
z_0\Delta \alpha(t) &
  \end{aligned} \right)
 &=&z_0
    \left( \begin{aligned}
\Delta \delta_0+\mu_\delta (t-t_0) & \\
\Delta \alpha_0+\mu_\alpha (t-t_0)&
  \end{aligned} \right)+
  a\bigg\{
    \left( \begin{aligned}
A & \\
B&
  \end{aligned} \right)X(t)+    \left( \begin{aligned}
F & \\
G&
  \end{aligned} \right)Y(t)\bigg\},~~~~~
  \label{eq:pos-mXt}
\end{eqnarray}
where $z_0$ is the distance to the binary system.

Similarly, we present the component along the line of sight that is complete with the fixed radial distance $\Delta z_0$ to the star and its radial-velocity $v_r $, as
{}
\begin{eqnarray}
\Delta z(t)&=& \Delta z_0+v_r (t-t_0)+a \Big(CX(t)+HY(t)\Big),
\label{eq:pos-mZt}
\end{eqnarray}
where $X(t)$ and $Y(t)$ are from (\ref{eq:pos-X}) and (\ref{eq:pos-Y})
{}
\begin{eqnarray}
X(t)&=&\cos E(t)-e,
\label{eq:pos-Xtt2}\\
Y(t)&=& \sqrt{1-e^2}\sin E(t),
\label{eq:pos-Ytt2}
\end{eqnarray}
and $E(t)$ is determined from Kepler's equation  (\ref{eq:pos-M1}) in the form of the following transcendent equation:
{}
\begin{eqnarray}
E-e\sin E= \frac{2\pi}{T}\big(t-t_0\big)\equiv n(t-t_0)=M,
\label{eq:pos-M2}
\end{eqnarray}
which for small eccentricities may be approximated as
{}
\begin{eqnarray}
E= M+e\sin M+{\textstyle\frac{1}{2}}e^2\sin 2M +{\cal O}(e^3), \qquad{\rm with}\qquad  M=\frac{2\pi}{T}\big(t-t_0\big).
\label{eq:pos-M3}
\end{eqnarray}

\subsection{Modeling velocities and accelerations}

Taking the time derivatives from each of the expressions in (\ref{eq:pos-Xt}),
with the help of (\ref{eq:pos-Xtt2})--(\ref{eq:pos-M2}),
we obtain velocities and accelerations in the corresponding directions. To do that we need only to take appropriate derivatives from $X$ and $Y$ given by (\ref{eq:pos-X}) and (\ref{eq:pos-Y}). To do that, we use (\ref{eq:pos-vec-E-dor}) and compute
{}
\begin{eqnarray}
\dot X&=&-\frac{n\sin E}{1-e\cos E}, \qquad\qquad \dot Y= \frac{n\sqrt{1-e^2}\cos E}{1-e\cos E},
\label{eq:pos-X-dot}\\
\ddot X&=& -\frac{n^2(\cos E-e)}{(1-e\cos E)^3}, \qquad ~~
\ddot Y= -\frac{n^2\sqrt{1-e^2}\sin E}{(1-e\cos E)^3}.
\label{eq:pos-Y-dot}
\end{eqnarray}

Thus,  taking the time derivatives from $x$ and $y$, we obtain the velocities $v_\delta=\dot x$ and $v_\alpha=\dot y$, in the ${\rm Dec}$ and ${\rm RA}$ directions respectively. Time derivative int the $z$ direction, $v_z=\dot z$ results in the radial velocity. Using (\ref{eq:pos-X-dot}), the corresponding expressions are given as
{}
\begin{eqnarray}
    \left( \begin{aligned}
v_\delta & \\
v_\alpha & \\
v_z & \\
  \end{aligned} \right) \equiv
   \left( \begin{aligned}
\dot x& \\
\dot y & \\
\dot z &
  \end{aligned} \right)
 &=&
  a\bigg\{
    \left( \begin{aligned}
A & \\
B & \\
C & \\
  \end{aligned} \right)\dot X(t)+    \left( \begin{aligned}
F & \\
G&  \\
H & \\
  \end{aligned} \right)\dot Y(t)\bigg\}=
 \frac{na}{1-e\cos E}
 \bigg\{
    -\left( \begin{aligned}
A & \\
B & \\
C & \\
  \end{aligned} \right)\sin E +    \left( \begin{aligned}
F & \\
G&  \\
H & \\
  \end{aligned} \right)\sqrt{1-e^2}\cos E \bigg\}.~~~~~
  \label{eq:pos-Xt-dot}
\end{eqnarray}

Similarly, with the help of (\ref{eq:pos-Y-dot}), we determine the accelerations
{}
\begin{eqnarray}
    \left( \begin{aligned}
a_\delta & \\
a_\alpha & \\
a_z & \\
  \end{aligned} \right) &\equiv&
   \left( \begin{aligned}
\ddot x& \\
\ddot y & \\
\ddot z &
  \end{aligned} \right)
 =
  a\bigg\{
    \left( \begin{aligned}
A & \\
B & \\
C & \\
  \end{aligned} \right)\ddot X(t)+    \left( \begin{aligned}
F & \\
G&  \\
H & \\
  \end{aligned} \right)\ddot Y(t)\bigg\}=\nonumber\\[-5pt]
  &=&
-\frac{n^2a}{(1-e\cos E)^3}
 \bigg\{
  \left( \begin{aligned}
A & \\
B & \\
C & \\
  \end{aligned} \right)\big(\cos E -e\big) +    \left( \begin{aligned}
F & \\
G&  \\
H & \\
  \end{aligned} \right)\sqrt{1-e^2}\sin E \bigg\}=
  -\omega^2(t)
   \left( \begin{aligned}
x & \\
y & \\
z & \\
  \end{aligned} \right),~~~~~
  \label{eq:pos-Xt-ddot}
\end{eqnarray}
where
{}
\begin{eqnarray}
\omega^2(t) =\frac{n^2}{(1-e\cos E)^3}=\Big(\frac{2\pi}{T}\Big)^2\frac{1}{(1-e\cos E)^3}.
\label{eq:omega-t}
\end{eqnarray}

These expressions allow the calculations of all the expressions that involve planetary velocities and accelerations.

We also note that each object in a binary system orbit the center of mass of the system in an ellipse with the same eccentricity but the semi-major axes is reduced in scale by a factor
{}
\begin{eqnarray}
a_1=\frac{m_2}{m_1+m_2}a \qquad {\rm and}\qquad a_2=-\frac{m_1}{m_1+m_2}a.
\label{eq:sem-maj}
\end{eqnarray}
The orbital periods of the two objects must each be equal to $T$ and therefore the two mean motions must also equal to be $(n_1=n_2=n)$, although the semi-major axes are not. Each mass them moves on its own elliptical orbit with respect to the common center of mass, and the periapses of their orbits differ by $\pi$.

\end{document}